\documentclass[%
reprint,
nofootinbib,
amsmath,amssymb,
aps,
prd,
superscriptaddress]{revtex4-2}

\usepackage{graphicx}
\usepackage{dcolumn}
\usepackage{bm}
\usepackage{dcolumn} 
\usepackage{hyperref}
\hypersetup{colorlinks=true,citecolor=blue,filecolor=blue,urlcolor=blue,linkcolor=blue}
\usepackage{orcidlink}
\usepackage{booktabs}

\usepackage{xcolor}

\usepackage{siunitx}

\renewcommand{\arraystretch}{1.4}

\DeclareSIUnit\kpc{kpc}
\DeclareSIUnit\Mpc{Mpc}
\DeclareSIUnit\Gpc{Gpc}
\DeclareSIUnit\Gyr{Gyr}

\DeclareSIUnit{\hub}{\mathit{h}}
\DeclareSIUnit{\invhub}{\mathit{h}^{-1}}
\newcommand{\SIF}[2]{\SI[parse-numbers=false]{#1}{#2}}

\usepackage{aas_macros}

\usepackage[nameinlink,noabbrev]{cleveref}
\crefname{equation}{Eq.}{Eqs.}
\crefname{section}{Section}{Sections}
\crefname{figure}{Fig.}{Figures}
\crefname{table}{Table}{Tables}
\crefname{appendix}{Appendix}{Appendices}
\Crefname{figure}{Figure}{Figures}
\Crefname{equation}{Equation}{Equations}
\Crefname{section}{Section}{Sections}
\Crefname{table}{Table}{Tables}

\newcommand{\sumnu}{\sum m_\nu}
\newcommand{\sumnueff}{\sum m_{\nu,{\rm eff}}}

\newcommand{\lcdm}{$\Lambda$CDM}

\begin{document}

\preprint{APS/123-QED}

\title{Constraints on Neutrino Physics from DESI DR2 BAO and DR1 Full Shape}


\author{W.~Elbers\orcidlink{0000-0002-2207-6108}}
\affiliation{Institute for Computational Cosmology, Department of Physics, Durham University, South Road, Durham DH1 3LE, UK}

\author{A.~Aviles\orcidlink{0000-0001-5998-3986}}
\affiliation{Instituto de Ciencias F\'{\i}sicas, Universidad Nacional Aut\'onoma de M\'exico, Av. Universidad s/n, Cuernavaca, Morelos, C.~P.~62210, M\'exico}
\affiliation{Instituto Avanzado de Cosmolog\'{\i}a A.~C., San Marcos 11 - Atenas 202. Magdalena Contreras. Ciudad de M\'{e}xico C.~P.~10720, M\'{e}xico}

\author{H.~E.~Noriega\orcidlink{0000-0002-3397-3998}}
\affiliation{Instituto de Ciencias F\'{\i}sicas, Universidad Nacional Aut\'onoma de M\'exico, Av. Universidad s/n, Cuernavaca, Morelos, C.~P.~62210, M\'exico}
\affiliation{Instituto de F\'{\i}sica, Universidad Nacional Aut\'{o}noma de M\'{e}xico,  Circuito de la Investigaci\'{o}n Cient\'{\i}fica, Ciudad Universitaria, Cd. de M\'{e}xico  C.~P.~04510,  M\'{e}xico}

\author{D.~Chebat\orcidlink{0009-0006-7300-6616}}
\affiliation{IRFU, CEA, Universit\'{e} Paris-Saclay, F-91191 Gif-sur-Yvette, France}

\author{A.~Menegas}
\affiliation{Institute for Computational Cosmology, Department of Physics, Durham University, South Road, Durham DH1 3LE, UK}

\author{C.~S.~Frenk\orcidlink{0000-0002-2338-716X}}
\affiliation{Institute for Computational Cosmology, Department of Physics, Durham University, South Road, Durham DH1 3LE, UK}

\author{C.~Garcia-Quintero\orcidlink{0000-0003-1481-4294}}
\affiliation{Center for Astrophysics $|$ Harvard \& Smithsonian, 60 Garden Street, Cambridge, MA 02138, USA}
\affiliation{NASA Einstein Fellow}

\author{D.~Gonzalez\orcidlink{0009-0009-6485-640X}}
\affiliation{Departamento de F\'{\i}sica, DCI-Campus Le\'{o}n, Universidad de Guanajuato, Loma del Bosque 103, Le\'{o}n, Guanajuato C.~P.~37150, M\'{e}xico.}

\author{M.~Ishak\orcidlink{0000-0002-6024-466X}}
\affiliation{Department of Physics, The University of Texas at Dallas, 800 W. Campbell Rd., Richardson, TX 75080, USA}

\author{O.~Lahav}
\affiliation{Department of Physics \& Astronomy, University College London, Gower Street, London, WC1E 6BT, UK}

% ...

\collaboration{DESI Collaboration}

\begin{abstract}
The Dark Energy Spectroscopic Instrument (DESI) Collaboration has obtained robust measurements of baryon acoustic oscillations (BAO) in the redshift range, $0.1 < z < 4.2$, based on the Lyman-$\alpha$ forest and galaxies from Data Release 2 (DR2). We combine these measurements with cosmic microwave background (CMB) data from \emph{Planck} and the Atacama Cosmology Telescope to place our tightest constraints yet on the sum of neutrino masses. Assuming the cosmological $\Lambda$CDM model and three degenerate neutrino states, we find $\sum m_\nu<\SI{0.0642}{\eV}$ (95\%) with a marginalized error of $\sigma(\sum m_{\nu})=\SI{0.020}{\eV}$. We also constrain the effective number of neutrino species, finding $N_\mathrm{eff}=3.23^{+0.35}_{-0.34}$ (95\%), in line with the Standard Model prediction. When accounting for neutrino oscillation constraints, we find a preference for the normal mass ordering and an upper limit on the lightest neutrino mass of $m_l < \SI{0.023}{\eV}$ (95\%). However, we determine using frequentist and Bayesian methods that our constraints are in tension with the lower limits derived from neutrino oscillations. Correcting for the physical boundary at zero mass, we report a 95\% Feldman-Cousins upper limit of $\sum m_\nu<\SI{0.053}{\eV}$, breaching the lower limit from neutrino oscillations. Considering a more general Bayesian analysis with an effective cosmological neutrino mass parameter, $\sum m_{\nu,\mathrm{eff}}$, that allows for negative energy densities and removes unsatisfactory prior weight effects, we derive constraints that are in $3\sigma$ tension with the same oscillation limit, while the error rises to $\sigma(\sum m_{\nu,\mathrm{eff}})=\SI{0.053}{\eV}$. In the absence of unknown systematics, this finding could be interpreted as a hint of new physics not necessarily related to neutrinos. The preference of DESI and CMB data for an evolving dark energy model offers one possible solution. In the $w_0w_a$CDM model, we find $\sum m_\nu<\SI{0.163}{\eV}$ (95\%), relaxing the neutrino tension. These constraints all rely on the effects of neutrinos on the cosmic expansion history. Using full-shape power spectrum measurements of Data Release 1 (DR1) galaxies, we place complementary constraints that rely on neutrino free streaming. Our strongest such limit in $\Lambda$CDM, using selected CMB priors, is $\sum m_\nu<\SI{0.193}{\eV}$ (95\%).
\end{abstract}

\maketitle

\tableofcontents

\section{Introduction}

The connection between cosmic neutrinos and the large-scale structure
of the Universe has been studied since the 1950s (e.g. \cite{Alpher:1953zz}), when the
concept of dark matter was beginning to take hold. One of the first
particles proposed as dark matter was a massive neutrino
created in the early Universe \cite{Gershtein_Zeldovich_66,Cowsik_McClelland_73,Szalay_76}. Years
later, a reported laboratory measurement of a mass of \SI{30}{\eV} for the
electron neutrino \cite{Lubimov_80} --a mass large enough to provide
the critical density needed to close the Universe-- gave great impetus
to the neutrino dark-matter model.

Such relatively light particles would remain relativistic until
relatively late times (hence their generic name `hot dark matter') and
their thermal motions would wash out density perturbations below a
free-streaming scale (which varies inversely with neutrino mass),
corresponding to the typical comoving distance that a particle travels
in the age of the Universe, $\sim \SI{40}{Mpc}$ for a \SI{30}{\eV} neutrino at
recombination \cite{Bond_80}. Shortly after, one of the first 
cosmological simulations showed that such a large damping scale would
result in a galaxy clustering pattern inconsistent with
observations \cite{Frenk_83,White_83}. That electron neutrinos could
not have a mass of $\sim \SI{30}{\eV}$, an early instance of input from
cosmology into particle physics, was later confirmed  experimentally \cite{Fritschi86,Wilkerson87,Belesev95,Weinheimer99,Lobashev99,Bonn01}.

The recent upper limit on the neutrino mass from the KATRIN experiment, combined with information from oscillation experiments, $\sum m_{\nu} < 1.35\,$eV (90\%) \cite{KATRIN19,KATRIN22,KATRIN24,Esteban:2024eli,nufit6,Capozzi_21,Salas_21}, indicates that neutrinos make up no more than a few percent of the total dark matter. However, their free streaming would have left an imprint in the power spectrum of density fluctuations, in the form of a small dip of amplitude $\Delta P(k)/P(k) \propto \Omega_\nu/\Omega_{\rm m}$ (where $\Omega$ is the mean density in units of the critical density), below the free-streaming scale. Such a distortion is, in principle, measurable in a sufficiently large and accurate galaxy survey. The first constraints on the sum of neutrino masses from the shape of the galaxy power spectrum were obtained from the 2dFGRS redshift survey, $\sum m_{\nu} \leq 1.8\,$eV (95\%) \cite{Elgaroy_02}, and subsequently from the SDSS \cite{Tegmark_04} in combination with cosmic microwave background (CMB) data from WMAP \cite{Bennett03,Spergel03}.

The free-streaming distortion can also be constrained by measuring the shape of the Lyman-$\alpha$ forest flux power spectrum along the line of sight, with the first such constraint, $\sum m_\nu<\SI{5.5}{\eV}$ (95\%), by \citet{Croft99} predating the galaxy power-spectrum limit from 2dFGRS. Significant improvements were made in subsequent years with the arrival of large samples of quasar spectra from SDSS and BOSS \cite{Seljak05,Goobar06,Seljak06,Viel10,PalanqueDelabrouille15,Rossi15,Yeche17}.

In addition to this small distortion of the fluctuation power
spectrum, neutrinos also affect the expansion history of the Universe
because they behave as radiation in the early Universe and as
dark matter at later times. Since the expansion
history of the Universe is reflected in the position of the peaks of 
the `baryon acoustic oscillations' (BAO) in the matter (and galaxy)
power spectra, BAO can also be used to constrain $\sum m_{\nu}$, when combined with other cosmological probes, particularly the power spectrum of temperature anisotropies of the CMB. In recent years, this signature has provided the strongest limits on $\sum m_\nu$ \cite{2012PhRvD..86j3518P,Vagnozzi17,Loureiro19,Choudhury20,PlanckCosmology2020,Palanque20,DiValentino21,Brieden22}, culminating in the latest constraints from the Dark Energy Spectroscopic Instrument (DESI) \cite{DESI2024.VI.KP7A,DESI2024.VII.KP7B,DESI.DR2.BAO.cosmo}.

Besides $\sum m_\nu$, the expansion history also depends on the number of neutrino species at the time when they are relativistic. The radiation density, in excess of that due to photons, can be expressed in terms of an effective number of neutrino species, $N_\mathrm{eff}$. In addition to the properties of cosmic neutrinos, measurements of $N_\mathrm{eff}$ constrain the existence of new neutrino-like relics that were in thermal equilibrium with the primordial plasma \cite{Gerbino18,Dvorkin22}. The Standard Model prediction is $N_\mathrm{eff}=3.044$ for the non-instantaneous decoupling of three neutrino species, incorporating corrections from flavour oscillations and finite temperature quantum electrodynamics \cite{Froustey20,Bennett21,Drewes24}.

DESI \cite{Snowmass2013.Levi,DESI2022.KP1.Instr,DESI2016b.Instr}, in combination with CMB data from \emph{Planck} \cite{PlanckLikelihood2020,PlanckLensing2022} and the Atacama Cosmology Telescope (ACT) \cite{ACTDR62024,ACTLensing2024}, has set the strongest current astrophysical constraints on $\sumnu$ \cite{DESI2024.VI.KP7A, DESI2024.VII.KP7B}. This has been achieved using two complementary approaches. One is based on the BAO traced by over 6 million galaxies, quasars, and the Lyman-$\alpha$ forest over the redshift range $0.1<z<4.2$ \cite{DESI2024.VI.KP7A}. The other is based on simultaneously analyzing the BAO and the power spectrum (`full shape', FS) of 4.7 million galaxies and quasars over the redshift range $0.1 < z < 2.1$ \cite{DESI2024.VII.KP7B}.  In both cases, in order to break parameter degeneracies, the DESI data can be supplemented with other data, such as the power spectrum of temperature anisotropies, polarization and lensing of the CMB. The estimates are always made within the framework of an assumed cosmological model.

Adding FS to the DESI BAO analysis, an interesting constraint can already be obtained without including CMB data, but adopting two external priors to break parameter degeneracies. The first is a prior on the physical baryon density, $\omega_\mathrm{b} \equiv \Omega_\mathrm{b} h^2$, with $h = H_0 / (100 \, \text{km} \, \text{s}^{-1} \text{Mpc}^{-1})$ the dimensionless Hubble constant, which can be determined from Big Bang Nucleosynthesis (BBN) \cite{Schoeneberg:2024}. The second is a prior on the spectral index of the matter power spectrum, $n_\mathrm{s}$, corresponding to ten times the uncertainty ($10\sigma$) from \emph{Planck} (denoted as $n_{\mathrm{s},10}$). Assuming the $\Lambda$CDM cosmology, the DESI upper limit for three degenerate neutrino species, from the combined FS and BAO datasets of Data Release 1 (DR1) was found to be \cite{DESI2024.VII.KP7B}
\begin{flalign}
\begin{aligned}
    &\qquad \text{DESI DR1 (FS+BAO) + BBN + $n_{\mathrm{s},10}$:} \\
    &\qquad \, \sumnu < 0.409 \,\si{\eV} \quad (95\%). 
\end{aligned}
&&
\end{flalign}

\noindent
This limit can be significantly tightened by adding CMB data, as was done in \cite{DESI2024.VII.KP7B}, yielding\footnote{The suffix (\texttt{plik}) is used to differentiate the baseline CMB likelihoods from \citep{PlanckLikelihood2020} used in \cite{DESI2024.VII.KP7B} from the baseline CMB dataset used in this paper (see \cref{sec:cmb_data}).}
\begin{flalign}
\begin{aligned}
    &\qquad \text{DESI\ DR1 (FS+BAO) + CMB (\texttt{plik}):} \\
    &\qquad \, \sumnu < 0.071 \,\si{\eV} \quad (95\%). 
\end{aligned}
&&
\end{flalign}

These upper limits were obtained adopting the minimal physical prior, $\sum m_\nu>0$, and assuming three degenerate neutrino mass states. The constraints are very
close to the lower limits from neutrino oscillations:  $\sumnu > \SI{0.059}{\eV}$ for the normal neutrino ordering\footnote{Normal ordering, or hierarchy, corresponds to the case when the smallest mass splitting among the three neutrino species is between the lowest mass eigenstates. In contrast, the inverted ordering refers to the case when the smallest mass splitting is between the two highest eigenstates.}, and $\sumnu > \SI{0.10}{\eV}$ for the inverted ordering \cite{Esteban:2024eli,nufit6,Capozzi_21,Salas_21}. Meanwhile, the DESI constraints on $N_\mathrm{eff}$ were always compatible with the Standard Model prediction \cite{DESI2024.VI.KP7A,DESI2024.VII.KP7B}.

One of the most tantalizing results from the analysis of the DESI DR1 data, when combined with CMB, is that the $\Lambda$CDM model is disfavored at $2.6\sigma$ in comparison to models with dynamical dark energy. This discrepancy increases to $3.9\sigma$ when DESY5 supernovae are also included \cite{DESI2024.VI.KP7A}. 
A similar conclusion was obtained from the DESI DR1 FS
analysis 
\cite{DESI2024.VII.KP7B} which, furthermore, showed
that the $\sum m_\nu$ constraints strongly depend on the assumed
dark energy model;
a consequence of the well-known degeneracy between $\sum m_\nu$ and the dark energy equation of state \cite{Hannestad05,Lorenz17,Vagnozzi18,Upadhye19,Liu20,Choudhury20,Sharma22,Choudhury24,Shao24,Yadav24}. 
The datasets are consistent with an evolving dark energy equation of state, $w=P/\rho$, based on constraints in terms of the CPL parametrization \cite{Chevallier:2001,Linder2003}, which assumes that the dark energy equation of state evolves with redshift, $z$, as
\begin{align}
w(z) = w_0 + w_a z / (1+z). \label{eq:w0wa_definition}
\end{align}

\noindent
For this `$w_0w_a$CDM' model, the same combination of data (DESI+CMB+DESY5)
relaxes the neutrino mass constraint to $\sum m_\nu <\SI{0.196}{\eV}$
(95\%), which is now consistent with neutrino oscillations for both mass orderings.

A puzzling aspect of the DESI analysis in both the BAO and FS+BAO cases is that the marginalized posterior distribution of $\sum m_\nu$ peaks at the lower edge of the prior, $\sum m_\nu=0$, and resembles the tail of a distribution with a central value in the negative mass range. Not only are the strongest constraints in slight tension with neutrino oscillations, presenting a close-to-$2\sigma$ discrepancy, but the posterior distributions are also dominated by ``prior weight effects'', which arise when the prior on a parameter forces the posterior away from its maximum likelihood value.

In fact, this feature was already present, albeit at less significance, in cosmological analyses prior to DESI. It was discussed by the \emph{Planck} \cite{Planck14} and SDSS \cite{Alam21} collaborations, but gained urgency following the release of the DESI DR1 results \cite{Craig24,Green24,Elbers_24,Allali24,Naredo-Tuero24,Noriega:2024lzo,Escudero24,Jiang25,Reboucas25}. Motivated by prior weight effects in these results, more general parametrizations were introduced to extend the neutrino mass analysis to negative values \cite{Craig24,Green24,Elbers_24}, finding that the tension with the oscillation limits increased to close to $3\sigma$ for $\Lambda$CDM. However, consistency could be restored in evolving dark energy models \cite{Elbers_24}.

We present here a new analysis using BAO measurements from the second data release (DR2) of DESI, which includes redshift data for over 14 million galaxies and quasars, as well as Lyman-$\alpha$ forest spectra from more than 820,000 quasars \cite{DESI.DR2.BAO.cosmo}. In \cref{sec:data_methods}, we present the data and methods used in our analysis. With these measurements, DESI is well placed to detect the distinctive imprints of massive neutrinos on cosmological observables, as discussed in \cref{sec:source} below. Our constraints on neutrino mass are the tightest obtained so far by any method and it is important to establish their robustness. We do this in \cref{sec:cmb_robustness}, where we introduce a new set of mock catalogs constructed from $N$-body simulations with a full treatment of neutrinos of different mass that we use for additional validation of our BAO methodology.

The baseline results of our analysis are then presented in \cref{Standard}.  In this paper, we combine the DESI data with constraints from neutrino oscillation experiments to set limits on the lightest neutrino mass and to determine the preference for the normal ordering  (\cref{sec:neutrino_mass_ordering}). The matter of the neutrino mass tension is addressed in \cref{sec:neutrino_tension}, using both frequentist and Bayesian approaches. In \cref{sec:effective_neutrinos}, we perform an analysis using the effective cosmological neutrino mass parameter, $\sum m_{\nu,\mathrm{eff}}$, of \cite{Elbers_24}.

Previous DESI analyses of neutrino mass left a number of important questions open. One of them is the
actual source of the neutrino mass signal in the full-shape analysis: geometry, expansion rate, the shape of the density perturbation power spectrum, or a combination of these. We discuss this question in \cref{sec:full_shape_sources}. Finally, our conclusions are presented in \cref{sec:conclusions}.

\section{Data and methodology}\label{sec:data_methods}

In this section we provide a brief overview of the DESI DR2 BAO and DR1 FS data and modeling, as well as the external datasets and codes used in the analysis.

DESI stands out as the current largest and most precise spectroscopic galaxy survey \cite{Snowmass2013.Levi,DESI2022.KP1.Instr,DESI2016b.Instr}. The instrument is equipped with 5,000 fibers \cite{FiberSystem.Poppett.2024} in a robotic focal plate assembly \cite{FocalPlane.Silber.2023} on the Mayall Telescope at Kitt Peak National Observatory. With a high-performing optical design based on a 3.2 degree prime-focus corrector \cite{Corrector.Miller.2023}, DESI plans to measure over 40 million galaxy redshifts during a five-year period \cite{DESI2016a.Science}. Indeed, with its target selection \cite{TS.Pipeline.Myers.2023} over the imaging Legacy Survey \cite {LS.Overview.Dey.2019, BASS.Zou.2017}, DESI had a successful survey validation campaign \cite{DESI2023a.KP1.SV} (including visual inspections \cite{VIGalaxies.Lan.2023,VIQSO.Alexander.2023}) with an early data release \cite{DESI2023b.KP1.EDR}. DESI survey operations \cite{SurveyOps.Schlafly.2023} already provided us with the first data release (DR1) \cite{DESI2024.I.DR1}, and measurements of galaxy clustering \cite{DESI2024.II.KP3} were used to derive precise BAO measurements, from both galaxy and quasar clustering \cite{DESI2024.III.KP4} and the Lyman-$\alpha$ forest  \cite{DESI2024.IV.KP6}, as well as a FS analysis of galaxies and quasars \cite{DESI2024.V.KP5}. DR1 cosmological results were presented in \cite{DESI2024.VI.KP7A,DESI2024.VII.KP7B}. The second data release (DR2), used in this analysis, includes three years of observations, which were processed with spectroscopic reduction \cite{Spectro.Pipeline.Guy.2023} and redshift estimation \cite{Redrock.Bailey.2024,RedrockQSO.Brodzeller.2023} pipelines.

\subsection{DESI DR2 BAO}\label{sec:DESI_DR2_BAOA}

We use BAO measurements from DESI DR2, based on 14 million galaxy and quasar spectroscopic redshift measurements, and 820,000 Lyman-$\alpha$ forest spectra, as well as their cross-correlation with 1.2 million quasar positions. DESI can measure the BAO scale along the line of sight and transverse to it, expressed either as the ratio of a line-of-sight comoving distance, $D_\mathrm{H}(z)$, or a transverse comoving distance, $D_\mathrm{M}(z)$, to the sound horizon at the drag epoch, $r_\mathrm{d}$. On the other hand, isotropic BAO measurements constrain $D_\mathrm{V}/r_\mathrm{d}$, where $D_\mathrm{V}=\sqrt[3]{zD_\mathrm{M}^2D_\mathrm{H}}$ is the spherically-averaged distance.

The data are split into BAO measurements at various effective redshifts given by the different combinations of DESI tracers: the bright galaxy survey (BGS) \cite{BGS.TS.Hahn.2023}, luminous red galaxies (LRG) \cite{LRG.TS.Zhou.2023}, emission line galaxies (ELG) \cite{ELG.TS.Raichoor.2023}, and quasars (QSO) with their Lyman-$\alpha$ forests \cite{QSO.TS.Chaussidon.2023}. In the case of our lowest redshift measurement, the BGS tracer in the range $0.1<z<0.4$, we use the isotropic BAO measurement of $D_\mathrm{V}/r_\mathrm{d}$. The rest of our BAO measurements from galaxy and quasar clustering consist of a combination of correlated measurements of $D_\mathrm{H}/r_\mathrm{d}$ and $D_\mathrm{M}/r_\mathrm{d}$, namely: anisotropic BAO measurements for the LRG tracer in the redshift bins $0.4<z<0.6$ and $0.6<z<0.8$; a combined tracer measurement for LRG+ELG in $0.8<z<1.1$; another ELG tracer at $1.1<z<1.6$; and the QSO tracer in the range $0.8<z<2.1$. These BAO measurements are described in detail in \cite{DESI.DR2.BAO.cosmo} and their systematic error tests are presented in \cite{Y3.clust-s1.Andrade.2025}. Additionally, we complement these studies with further tests in the presence of massive neutrinos, as discussed in \cref{sec:cmb_robustness}. Finally, we also include BAO measurements of $D_\mathrm{H}/r_\mathrm{d}$ and $D_\mathrm{M}/r_\mathrm{d}$ from the Lyman-$\alpha$ forest auto-correlation and cross-correlation with the QSOs, covering a higher redshift range $1.8<z<4.2$; see \cite{DESI.DR2.BAO.lya} for the measurement details, and their companion papers~\cite{Y3.lya-s1.Casas.2025,Y3.lya-s2.Brodzeller.2025}. 

\subsection{DESI DR1 Full Shape}\label{sec:DESIDR!FS}

We use the Effective Field Theory of Large Scale Structure (EFT) \cite{McDonald2009,2012JCAP...07..051B,2012JHEP...09..082C,Vlah2015} to model the power spectrum multipoles of DR1 galaxies and quasars in a full-shape (FS) analysis. To this end, we rely on the same pipeline used for the DR1 publications 
\cite{DESI2024.V.KP5, DESI2024.VII.KP7B}, but allow the neutrino mass to vary, as explored in \cite{KP5s3-Noriega}. We use the monopole and quadrupole moments of the power spectrum across 6 redshift bins: BGS ($0.1<z<0.4$), LRG1 ($0.4<z<0.6$), LRG2 ($0.6<z<0.8$), LRG3 ($0.8<z<1.1$), ELG1 ($1.1<z<1.6$), QSO ($0.8<z<2.1$). For further details on this analysis, including blinding strategies, random catalogs, the window function, radial and angular integral constraints, fiber-assignment mitigation, and the combination of galactic caps, the reader is referred to \cite{DESI2024.V.KP5} and the supporting DESI references therein. Additionally, these sources provide broader discussions on the effects of various systematics arising from theoretical modeling \cite{KP5s1-Maus,KP5s2-Maus,KP5s3-Noriega,KP5s4-Lai,KP5s5-Ramirez}, halo occupation distribution (HOD) assumptions \cite{KP5s7-Findlay}, fiducial cosmology \cite{KP5s8-Findlay}, and both imaging and spectroscopic considerations \cite{KP5s6-Zhao}.

The baseline fitting range of scales is $0.02 < k \,/ (h \,\text{Mpc}^{-1}) < 0.20$, covering the regime where nonlinear physics is still under control within EFT. Apart from the cosmological parameters, the modeling includes 7 nuisance parameters per tracer, that account for nonlinear contributions through EFT, the bias prescription, and stochastic effects. Some of these parameters are degenerate with those specifying the cosmological model, potentially leading to projection effects. To partially mitigate these effects, we implement physically motivated priors on the nuisance parameters; see \cite{DESI2024.V.KP5}.

Accurate covariance matrices are essential for both the BAO and FS analyses. For DR1, these were computed from 1000 samples of EZmocks, with corrections based on studies of DESI DR1 mocks from the \texttt{Abacus} suite and analytical Gaussian methods using the RascalC algorithm \cite{2023MNRAS.524.3894R, KP4s7-Rashkovetskyi, KP4s6-Forero-Sanchez}.

As expected, the FS description naturally incorporates information from the BAO peak, leading to a strong cross-correlation that must be accurately modeled to combine both analyses. In particular, the covariances derived from the calibrated EZmocks effectively captured these cross-correlations for the DR1 results presented in \cite{DESI2024.V.KP5, DESI2024.VII.KP7B}.

In this paper, we present constraints from DESI DR1 (FS+BAO) that properly account for the cross-correlation between the FS and BAO measurements of DR1 galaxies. As the DR2 FS analysis is still blinded, the combination of DR1 FS and DR2 BAO provides the most powerful dataset available. We estimated that this combination provides a slight improvement over the constraints that are possible with DESI DR2 BAO alone. However, a proper calculation of the cross-correlation between DR1 FS and DR2 BAO requires mock catalogs that are not yet available. We therefore do not present analyses of this combination.

\subsection{CMB data}\label{sec:cmb_data}

We additionally combine our measurements with various external CMB datasets. We utilize measurements of temperature and polarization anisotropies from \emph{Planck} \cite{PlanckLikelihood2020} and lensing measurements from both \emph{Planck} \cite{PlanckLensing2022} and the Atacama Cosmology Telescope (ACT) \cite{ACTDR62024}. The likelihoods used in this paper include the original released likelihoods from the \emph{Planck} collaboration \citep{PlanckLikelihood2020} and further releases made by \citet{Rosenberg22} and \citet{Tristram2021,Tristram2024} in the years since. 

The CMB likelihoods are broadly broken down into two components, the first, computed for high-$\ell$ (i.e. $\ell > 30$) components of the temperature and polarization auto and cross spectra; the second, computed for low-$\ell$ (i.e. $\ell \leq 30$) temperature and polarization auto spectra. The two regimes are treated differently due to a breakdown in the Gaussian approximations that can be made at high $\ell$. Anisotropies at low $\ell$ must be treated differently due to the non-Gaussian and highly correlated measurements.

We use the following combinations of likelihoods: 
\begin{itemize}
    \item PR3 \texttt{plik}: The original likelihoods from \citep{PlanckLikelihood2020}, where we specifically use the \texttt{SimAll} and \texttt{Commander} likelihoods for the low-$\ell$ temperature and polarization auto spectra (low-$\ell$ TTEE) and the \texttt{plik} likelihoods for the high-$\ell$ temperature and polarization auto and cross spectra (high-$\ell$ TTTEEE).
    \item PR4 \texttt{CamSpec}: The high-$\ell$ TTTEEE likelihoods are replaced with the \texttt{CamSpec} likelihood \cite{Rosenberg22,Efstathiou2021}. The low-$\ell$ TTEE remain the same as the likelihoods for \texttt{plik} PR3 above.
    \item PR4 \texttt{LoLLiPoP-HiLLiPoP} or \texttt{L-H}: Both the low-$\ell$ TTEE and high-$\ell$ likelihoods are replaced with the \texttt{LoLLiPoP} and \texttt{HiLLiPoP} likelihoods of \citet{Tristram2021,Tristram2024}.
\end{itemize}
Finally, we make use of a CMB lensing likelihood based on a combination of \emph{Planck} and ACT DR6 data \cite{ACTDR62024, ACTLensing2024, PlanckLensing2022}. 

The baseline CMB dataset used in this paper, denoted simply as `CMB' from now on, consists of low-$\ell$ TTEE from \texttt{Commander} and \texttt{SimAll}, high-$\ell$ TTTEEE from \texttt{CamSpec}, and CMB lensing as described above.

\subsection{SNe data}

In this analysis we also use distance measurements from three compilations of Type Ia supernovae (SNe): Pantheon+ \citep{PantheonPlus2022,Brout:2022}, Union3 \citep{Union32023} and DESY5 \citep{DESY5SN2024}. Type Ia SNe have been observed since the late 90s from various survey programs utilizing many different telescopes in different observing conditions. Pantheon+ and Union3 both attempt to compile various SN observations from public and private legacy data, with much of these datasets overlapping between the two compilations. DESY5 constitutes an independent set of SNe data obtained from photometric observations of the Dark Energy Survey (DES) for $z > 0.1$, while also incorporating supernovae from spectroscopic surveys at low redshits, $z<0.1$, including some from Union3 and Pantheon+.

{
\renewcommand{\arraystretch}{1}
\begin{table}[t] 
    \centering
    \begin{tabular}{|llll|}
    \hline
    parametrization & parameter & default & prior\\  
    \hline 
    $\mathbf{\Lambda}$\textbf{CDM} & $\omega_\mathrm{cdm}$ &---& $\mathcal{U}[0.001, 0.99]$ \\   
    & $\omega_\mathrm{b}$ &---& $\mathcal{U}[0.005, 0.1]$ \\
    & $100\theta_\mathrm{MC}$\footnote{In some cases, where we use the \texttt{CLASS} code, we adopt $100\theta_*$ instead of $100\theta_\mathrm{MC}$. This has no effect on the results.} &---& $\mathcal{U}[0.5, 10]$ \\
    & $\ln(10^{10} A_\mathrm{s})$ &---& $\mathcal{U}[1.61, 3.91]$ \\
    & $n_\mathrm{s}$ &---& $\mathcal{U}[0.8, 1.2]$ \\
    & $\tau$ &---& $\mathcal{U}[0.01, 0.8]$ \\
    \hline 
    \textbf{Dark energy} & $w_0$ or $w$ & $-1$ & $\mathcal{U}[-3, 1]$ \\
    & $w_{a}$ & $0$ & $\mathcal{U}[-3, 2]$ \\
    \hline
    \textbf{Baseline neutrinos} & $\sumnu \; [\,\si{\eV}\,]$ & $0.06$ & $\mathcal{U}[0, 5]$ \\
    & $N_\mathrm{eff}$ & $3.044$ & $\mathcal{U}[0.05, 10]$  \\
    \textbf{Ordered neutrino} & $m_l \; [\,\si{\eV}\,]$ & --- & $\mathcal{U}[0, 5]$ \\
    \textbf{masses} & $\Delta m^2_{21} \; [\,\si{\eV^2}\,]$ & --- & $\mathcal{N}[\mu_{21},\sigma^2_{21}]$\footnote{See \cref{eq:solar_split,eq:atm_split} for the values of $\mu_{ab}$ and $\sigma^2_{ab}$.} \\
     & $\Delta m^2_{31} \; [\,\si{\eV^2}\,]$ & --- & $\mathcal{N}[\mu_{31},\sigma^2_{31}]$ \\
     & $\Delta m^2_{32} \; [\,\si{\eV^2}\,]$ & --- & $\mathcal{N}[\mu_{32},\sigma^2_{32}]$ \\
     & $\mathcal{M}$ & --- & $\mathcal{U}[-1, 1]$ \\
    \textbf{Effective neutrino} & $\sum m_{\nu,\mathrm{eff}}$ $[\,\si{\eV}\,]$ & $0.06$ & $\mathcal{U}[-5, 5]$ \\
    \textbf{masses} & & & \\
    \hline
    \end{tabular}
    \caption{
    Parameters and priors used in the analysis. Here $\mathcal{U}[{\rm min, max}]$ denotes a uniform prior over the specified range, while $\mathcal{N}(\mu, \sigma^2)$ represents a Gaussian prior with mean $\mu$ and standard deviation $\sigma$. In addition to the flat priors on $w_0$ and $w_a$ listed in the table, we also impose the requirement that $w_0+w_a<0$ in order to enforce a period of high-redshift matter domination.
    }
    \label{tab:priors}
\end{table}
}

\subsection{Neutrino oscillation data}

In our analysis, we also rely on terrestrial constraints on the neutrino mass splittings, obtained from a global fit to solar, atmospheric, reactor, and accelerator experiments \cite{Esteban:2024eli,nufit6,Capozzi_21,Salas_21}.
We specifically use the values for the mass squared differences from NuFIT 6.0 \cite{Esteban:2024eli,nufit6}, 
\begin{align}
\Delta m_{21}^2&=7.49\pm 0.19\times 10^{-5}\text{ eV}^2, \label{eq:solar_split}\\
\Delta m_{3\ell}^2&=\begin{cases}
+2.513\pm0.020\times 10^{-3}\text{ eV}^2 \quad \text{ (NO)},\\
-2.484\pm0.020\times 10^{-3}\text{ eV}^2 \quad \text{ (IO)},\end{cases} \label{eq:atm_split}
\end{align}

\noindent
with $\ell=1$ for the normal ordering (NO) and $\ell=2$ for the inverted ordering (IO). These imply lower limits on the sum of neutrino masses,
\begin{align}
    \sum m_{\nu} > \begin{cases}
    0.05878\pm 0.00023\,\si{\eV} & \text{(NO)},\\
    0.09892\pm 0.00041\,\si{\eV} & \text{(IO)}.
    \end{cases}
\end{align}

\noindent
For our purposes in this paper, the uncertainties on these lower limits can be neglected.

\subsection{Cosmological inference}\label{sec:cosmological_inference}

We use the \texttt{Cobaya} code \cite{Torrado:2021,Torrado:2019} for cosmological parameter inference employing Metropolis-Hastings Markov-Chain-Monte-Carlo (MCMC) sampling. \texttt{Cobaya} in turn calls the Boltzmann codes \texttt{CAMB} \cite{LewisCamb1999} and \texttt{CLASS} \cite{Lesgourgues11,Lesgourgues11b}, as well as likelihoods for various datasets. The posterior chains are analyzed to derive summary statistics and plots using the \texttt{getdist} package \cite{Lewis:2019xzd}.

The first set of runs are performed with a \lcdm\ model, specified by the first 6 parameters listed in \cref{tab:priors},
complemented with the sum of neutrino masses. The parameter $\theta_\text{MC}$ is an approximation to the acoustic angular scale $\theta_\ast$. This parameter is used only to make steps in the MCMC algorithm, while the actual likelihood calculations employ $\theta_*$. The remaining parameters are the the amplitude of primordial scalar perturbations, $A_\mathrm{s}$, their spectral index, $n_\mathrm{s}$, and the optical depth, $\tau$. The same model is later considered but with the sum of neutrino masses replaced by the effective number of relativistic degrees of freedom, $N_\mathrm{eff}$, as a varied parameter.  

We also consider variations of this model that have an additional parameter to represent a constant equation of state of dark energy that is allowed to take values different from $w=-1$, denoted as ${w}${CDM+}${\sum m_\nu}$. A further extension is considered where the equation of state, $w(z)$, is allowed to vary in time with the parameter $w_0$ being its value today and $w_a$ the slope of its time evolution, denoted as ${w_0w_a}${CDM+}${\sum m_\nu}$. Prior ranges on these cosmological parameters are all given in \cref{tab:priors}.

In subsequent analyses, the neutrino masses are either modeled using a parametrization in terms of the lightest neutrino mass, $m_l$, and squared mass splittings, or in terms of an effective neutrino mass parameter, $\sum m_{\nu,\mathrm{eff}}$. Details of these prescriptions are given in the corresponding sections. Most analyses use the \texttt{CAMB} code. In some cases, when considering the $\sum m_{\nu,\mathrm{eff}}$ parameter, we instead use \texttt{CLASS}.

\subsection{External priors}\label{sec:priors}

We also make use of certain external priors that compress the information from other experiments on parameters that are not measured well by DESI alone. These include a Gaussian prior on $\omega_\mathrm{b}$ from Big Bang Nucleosynthesis (BBN) \cite{Schoeneberg:2024}, given by $\omega_\mathrm{b}=0.02218\pm0.00055$. Following \cite{DESI2024.VII.KP7B}, we also use a Gaussian prior on $n_\mathrm{s}$, centered on the \emph{Planck} value, $n_\mathrm{s}=0.9649\pm0.0042$ \cite{PlanckCosmology2020}. In many cases, we adopt a looser prior with 10 times this uncertainty, $n_\mathrm{s}=0.9649\pm0.042$, which we denote as $n_{\mathrm{s},10}$. Finally, we also use more expansive multivariate Gaussian priors that compress the CMB information on $(\theta_*,\omega_\mathrm{b},\omega_\mathrm{cdm})$ or $(\theta_*,n_\mathrm{s})$. We constructed these priors by analyzing the baseline CMB dataset (including lensing) for a $\Lambda$CDM model with the effective neutrino mass parameter, $\sum m_{\nu,\mathrm{eff}}$, to properly capture the broader uncertainties that apply in this more general case.

\subsection{Validation on mocks}\label{sec:cmb_robustness}

Here, we present the results of an additional validation test of the neutrino analysis using a new set of mock catalogs that target massive neutrinos. Although the methodology has been thoroughly tested with mocks before (see \cite{DESI2024.III.KP4,Y3.clust-s1.Andrade.2025,KP4s9-Perez-Fernandez,KP4s10-Mena-Fernandez, KP4s11-Garcia-Quintero} and many references therein), including with mocks based on high-fidelity simulations from the \texttt{AbacusSummit} suite \cite{Garrison19,Abacus21}, we corroborate these results with a new set of high-fidelity simulations from the Peregrinus suite \cite{peregrinus25}, which were specifically designed for neutrino mass analyses. Key differences with high-fidelity simulations used previously by DESI include the use of third-order initial conditions, the choice of $N$-body code, the method for identifying bound structures, and the implementation of massive neutrino particles.

Let us briefly summarize the key properties of the Peregrinus simulations. They were carried out with the cosmological $N$-body code SWIFT \cite{schaller18,schaller24}, using the particle mesh and fast multipole methods for the gravitational force calculations, on the COSMA-8 system in Durham. The initial conditions were generated with third-order Lagrangian perturbation theory at $z=31$ using the monofonIC code \cite{hahn20,michaux20}, accounting for the presence of neutrinos \cite{elbers22,elbers22b}. The simulations followed the evolution of $6000^3$ dark matter and baryon particles and $3000^3$ massive neutrino particles in a comoving volume of $(\SIF{3h^{-1}}{\Gpc})^3$, giving a dark matter particle mass of $m_p=\SIF{10^{10}h^{-1}}{M_\odot}$. Massive neutrinos were implemented in the simulation with the $\delta f$ method \cite{elbers20} to model efficiently their non-linear evolution. Dark matter structures were identified using the hierarchical bound tracing (HBT+) method \cite{han12,Han_18,forouhar25}, which is particularly suited for accurately identifying and tracking substructures.

From these simulations, we created a set of specialized mock catalogs that target the DESI Bright Galaxy Survey (BGS), using the halo occupation distribution (HOD) mock pipeline outlined in \cite{smith24}, adapted for use with the Peregrinus simulations. This process entails fitting an HOD to each of a number of galaxy luminosity bins ranging from magnitude $-17$ to $-22$, such that the resulting set of galaxies in each bin reproduces a target projected two-point correlation function and the entire set reproduces a target luminosity function. These target functions are designed to match the results of the DESI One-Percent survey \cite{DESI2023a.KP1.SV}. The pipeline performs this HOD-fitting procedure for each simulation in the suite, and produces a mock catalog using this HOD applied to the simulation snapshots or lightcones. 

\begin{figure}
    \centering
    \resizebox{\linewidth}{!}{
        \includegraphics[width=0.9\linewidth]{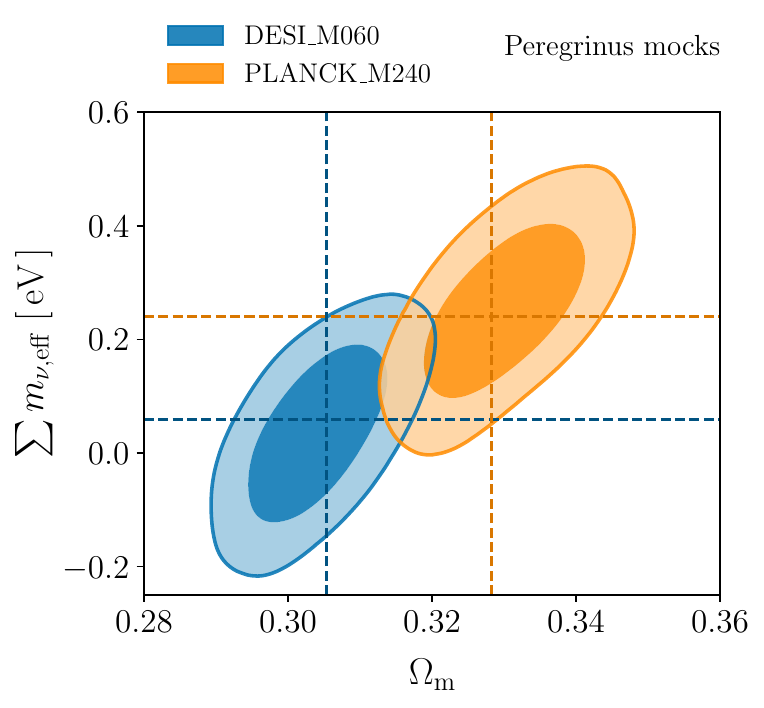}
    }
    \caption{Constraints on $\Omega_\mathrm{m}$ and $\sum m_{\nu,\mathrm{eff}}$ obtained from a BAO analysis of two large Peregrinus mocks, combined with a correlated CMB prior on the parameters $(\theta_*,\omega_\mathrm{cdm},\omega_\mathrm{b})$. The DESI\_M060 simulation has a DESI-like cosmology with $\sum m_\nu=\SI{0.059}{\eV}$ \cite{DESI2024.VI.KP7A} and the PLANCK\_M240 simulation has a \emph{Planck}-based cosmology with $\sum m_\nu=\SI{0.24}{\eV}$, the largest mass allowed by \emph{Planck} at 95\% \cite{PlanckCosmology2020}. The dashed lines indicate the true values in the simulations. The contours enclose 68\% and 95\% of the posterior volume.}
    \label{fig:mock_results}
\end{figure}

In the validation test presented here, we make use of two Peregrinus simulations. The first, labelled PLANCK\_M240, has a \emph{Planck}-based $\Lambda$CDM cosmology with a large neutrino mass of $\sum m_\nu=\SI{0.24}{\eV}$ \cite{PlanckCosmology2020}. This is the largest mass allowed at 95\% confidence by \emph{Planck} data alone. The second, labelled DESI\_M060, has a DESI-like $\Lambda$CDM cosmology \cite{DESI2024.VI.KP7A} with a neutrino mass of $\sum m_\nu=\SI{0.059}{\eV}$, the minimal value allowed by neutrino oscillations under the normal ordering. These values were chosen to obtain a reasonable range in $\sum m_\nu$ for our tests. The tests are based on cubic-box mocks at redshift $z=0.3$ with an absolute magnitude cutoff of $M_r<-21.35$, matching the fiducial DR2 cutoff, and RSD imposed along one of the Cartesian axes. We compute the galaxy correlation function from the pre-reconstruction distribution of galaxies and fit an isotropic BAO template, following the choices of \cite{DESI2024.III.KP4}. The resulting determination of the distance ratio, $D_\mathrm{V}/r_\mathrm{d}$, is combined with a synthetic correlated CMB prior on $(\theta_*,\omega_\mathrm{cdm},\omega_\mathrm{b})$, with the uncertainties matching the real baseline CMB dataset used in this paper, in an MCMC analysis to place constraints on the cosmological parameters. \cref{fig:mock_results} shows that we obtain unbiased constraints on $\Omega_\mathrm{m}$ and the effective sum of neutrino masses, $\sum m_{\nu,\mathrm{eff}}$, for both Peregrinus mocks. This confirms the expectation that the baseline DESI choices can be reliably applied in models with large neutrino masses, even when accounting for any possible non-linear effects or changes in galaxy-halo connection.

\section{Neutrinos in cosmology}\label{sec:source}

Neutrinos leave distinctive imprints on cosmological observables that DESI is uniquely positioned to detect. See the reviews \cite{Lesgourgues06,Wong11,Abazajian16} and references therein, as well as \cite{Loverde24,Noriega:2024lzo,Bertolez24,Escudero24,Racco25} for other recent studies on the source of the neutrino mass signal in cosmology.

Weak interactions in the early Universe produce a background of cosmic neutrinos, which decouple from the primordial plasma at a temperature of about $\SI{1}{\MeV}$. They subsequently move freely, essentially without interacting, and thus preserve their thermal phase-space distribution until late times.

However, their relativistic energies are lost over time due to cosmological redshift.  As the Universe expands, neutrinos transition from relativistic to non-relativistic states, with their energy density given by \cref{eq:neutrino_energy_density}. Neutrinos become non-relativistic when $m_\nu\gg T_\nu$, which happens at $z_\mathrm{nr}\approx 1890 \, (m_\nu/\si{\eV})$. In the relativistic limit, $z \gg z_\mathrm{nr}$, the energy density of neutrinos depends only on their number and temperature:
\begin{align}
    \rho_\nu(z \gg z_\mathrm{nr}) &= N_\mathrm{eff} \frac{7\pi^2}{120}T_{\nu,0}^4 (1+z)^4, \label{eq:Neff_def}
\end{align}

\noindent
where $N_\mathrm{eff}$ is the effective number of relativistic species and $T_{\nu,0}$ their present-day temperature. After becoming non-relativistic, $z \ll z_\mathrm{nr}$, their energy density scales with mass:
\begin{align}
    \rho_\nu(z \ll z_\mathrm{nr}) &= \frac{\sum m_\nu}{\SI{93.14}{\eV} h^2} \frac{3H_0^2}{8\pi G} (1+z)^3. \label{eq:rhonu_late}
\end{align}

\noindent
Hence, neutrinos contribute to the radiation density at early times and to the matter density at late times, leaving a unique imprint on the cosmic expansion history.

\subsection{Expansion history}\label{sec:expansion_history}

Galaxy BAO measurements determine a particular angular scale and a redshift separation,
\begin{align}
    \Delta\theta = \frac{r_\mathrm{d}}{D_\mathrm{M}(z)} \quad\quad\mathrm{and}\quad\quad \Delta z = \frac{r_\mathrm{d}}{c/H(z)}, \label{eq:bao}
\end{align}

\noindent
in the clustering of galaxies. These measurements probe the dimensionless ratio of a characteristic length scale, $r_\mathrm{d}$, imprinted on the matter distribution in the early Universe, to the late-time comoving angular diameter distance, $D_\mathrm{M}(z)$, or expansion length, $c/H(z)$, at the effective redshift, $z$, of the galaxy sample. The scale $r_\mathrm{d}=r_\mathrm{s}(z_\mathrm{d})$ corresponds to the sound horizon,
\begin{align}
    r_\mathrm{s}(z) = \int_{z}^\infty\frac{c_\mathrm{s}(z')}{H(z')}\mathrm{d}z',
\end{align}

\noindent
at the baryon drag epoch, $z_\mathrm{d}\approx1060$, when baryons decouple from photons. In the above expression, $c_\mathrm{s}$ is the sound speed of the primordial baryon-photon plasma. When it comes to neutrino masses, the power of the BAO technique lies in a key property that distinguishes neutrinos from cold dark matter: they are relativistic at the time, $z_\mathrm{d}$, when the scale $r_\mathrm{d}$ is imprinted in the matter distribution, but non-relativistic at the time when the BAO measurements are made.
 
The late-time quantities probed by BAO are the Hubble rate, $H(z)=H_0 E(z)$, and the transverse comoving distance, $D_\mathrm{M}(z)$. The former is given by
\begin{align}
    E(z) = \Big[&\Omega_\mathrm{cb}(1+z)^3 + \Omega_\mathrm{r} (1+z)^4 + 
                             \Omega_\mathrm{K}(1+z)^2 \;+  \label{eq:hubble_rate}\\
                              &\Omega_\nu \frac{\rho_\nu(z)}{\rho_{\nu,0}} + 
                              \Omega_\mathrm{DE}\frac{\rho_\mathrm{DE}(z)}{\rho_{\mathrm{DE},0}} \Big]^{1/2}. \nonumber
\end{align}

\begin{figure}
    \centering
    \resizebox{\linewidth}{!}{
        \includegraphics[width=0.95\linewidth]{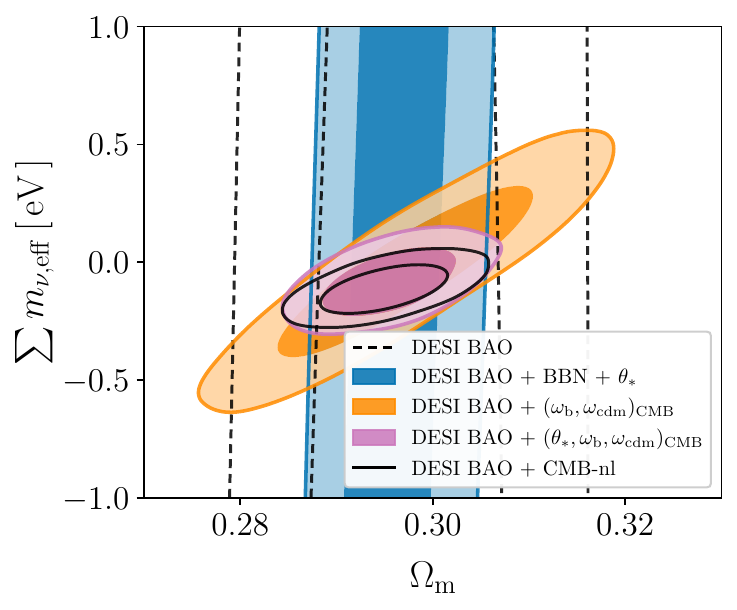}
    }
    \caption{Constraints on the sum of effective neutrino masses, $\sum m_{\nu,\mathrm{eff}}$, and $\Omega_\mathrm{m}$ for DESI DR2 BAO combined with different BBN and CMB priors (`nl' stands for no CMB lensing). The contours enclose 68\% and 95\% of the posterior volume. The figure demonstrates that BAO alone do not constrain $\sum m_{\nu,\mathrm{eff}}$. The addition of BBN and geometric CMB information, in the form of a prior on $\theta_*$, only improves the constraint on $\Omega_\mathrm{m}$. CMB priors on $\omega_\mathrm{b}$ and $\omega_\mathrm{cdm}$ help to break the degeneracy between the CDM and neutrino densities. When BAO is combined with a prior on $(\theta_*,\omega_\mathrm{b},\omega_\mathrm{cdm})$, the posterior approaches that of the full DESI BAO + CMB-nl combination.
    }
    \label{fig:sources_bao_cmb}
\end{figure}

\noindent
Here, $\Omega_\mathrm{cb}$ is the fraction of the critical density in cold dark matter and baryons, $\Omega_\mathrm{r}$ the fraction of radiation, $\Omega_\mathrm{K}$ the fraction of curvature, $\Omega_\nu$ the present fraction of massive neutrinos, and $\Omega_\mathrm{DE}=1-\Omega_\mathrm{cb}-\Omega_\mathrm{r}-\Omega_\mathrm{K}-\Omega_\nu$ the fraction of dark energy. In the $w_0w_a$CDM model, the dark energy density can be expressed as
\begin{align}
    \frac{\rho_\mathrm{DE}(z)}{\rho_{\mathrm{DE},0}} = (1+z)^{3(1+w_0+w_a)e^{-3w_az/(1+z)}}, \label{eq:rho_DE}
\end{align}

\noindent
which simplifies to a cosmological constant in the case that $w_0=-1$ and $w_a=0$. In a flat FLRW cosmology, the transverse comoving distance is
\begin{align}
    D_\mathrm{M}(z) = \frac{c}{H_0} \int_0^z\frac{ \mathrm{d}z'}{E(z')}. \label{eq:DM}
\end{align}

\noindent
DESI measures the BAO scale after the non-relativistic transition, $z\ll z_\mathrm{nr}$, when massive neutrinos contribute simply to the matter density, $\Omega_\mathrm{m}=\Omega_\mathrm{cb}+\Omega_\nu$. At late times, neglecting also $\Omega_\mathrm{r}$ and assuming a cosmological constant for simplicity, we thus have
\begin{align}
     E(z) \approx \sqrt{\Omega_\mathrm{m}(1+z)^3 + \Omega_\mathrm{DE}}.\label{eq:late_hubble}
\end{align} 

\noindent
Regardless of any calibration of $r_\mathrm{d}$, BAO measurements at different redshifts constrain the relative distances $D_\mathrm{M}(z_1)/D_\mathrm{M}(z_2)$ and expansion rates $E(z_1)/E(z_2)$. From \cref{eq:bao,eq:DM,eq:late_hubble}, we see that these measurements constrain $\Omega_\mathrm{m} = \Omega_\mathrm{cb} + \Omega_\nu$, but cannot distinguish massive neutrinos from cold dark matter. Moreover, \cref{eq:rhonu_late} shows that we additionally need to determine $h$ in order to infer $\sum m_\nu$ from $\Omega_\nu$, and this requires calibration of the standard ruler, $r_\mathrm{d}$.

To illustrate that $\sum m_\nu$ remains entirely unconstrained from BAO alone, we present the cosmological constraints from DESI DR2 BAO on $\Omega_\mathrm{m}$ and $\sum m_\nu$ as vertical dashed lines in \cref{fig:sources_bao_cmb}. These results were obtained using the data and methods described in detail in \cref{sec:data_methods}.

Additional information on $\Omega_\mathrm{m}$ can be obtained from measurements of the acoustic angular scale, $\theta_*=r_*/D_\mathrm{M}(z_*)$, in the CMB. Although this is a close analogue of the BAO feature, $\theta_*$ measures the sound horizon at the epoch of recombination, $r_*=r_\mathrm{s}(z_*)$. The relative timing of $z_*$ and $z_\mathrm{d}$ depends on the baryon density, which cannot be determined from BAO and $\theta_*$ alone. An external BBN prior \cite{Schoeneberg:2024} on the baryon density, $\omega_\mathrm{b}$, allows us to relate $r_*$ and $r_\mathrm{d}$, effectively extending the relative BAO measurements to $z_*$. The vertical blue band in \cref{fig:sources_bao_cmb} shows that this combination of DESI BAO + BBN + $\theta_*$ improves the precision on $\Omega_\mathrm{m}$, but is unable to constrain $\sum m_\nu$ as $\Omega_\nu$ and $\Omega_\mathrm{cb}$ remain degenerate.

Independent of $\theta_*$, CMB data constrain the densities of baryons, $\omega_\mathrm{b}$, and cold dark matter, $\omega_\mathrm{cdm}$, through the relative amplitudes of the acoustic peaks. As neutrinos are still relativistic at recombination, these amplitudes depend only very weakly on $\sum m_\nu$. This distinguishing feature makes it possible both to break the degeneracy between $\Omega_\mathrm{cb}$ and $\Omega_\nu$ and to calibrate the standard ruler, $r_\mathrm{d}$, thus providing the necessary information to constrain $\sum m_\nu$. This is shown by the orange contours in \cref{fig:sources_bao_cmb}.\footnote{Note that in the cases where we include priors on multiple CMB parameters, we account for their correlations; see \cref{sec:priors}.} We note here in passing that the shape of the galaxy power spectrum provides another way to distinguish between neutrinos and cold dark matter, as we will discuss in \cref{sec:full_shape_sources}.

The positive correlation between $\sum m_\nu$ and $\Omega_\mathrm{m}$, shown by the orange contours in \cref{fig:sources_bao_cmb}, can be understood as follows. \cref{eq:bao,eq:DM,eq:late_hubble} show that BAO measurements constrain both $H_0r_\mathrm{d}$, from the overall amplitude of the distance measurements, and $\Omega_\mathrm{m}$, from their redshift dependence, resulting in a tight positive correlation between $\Omega_\mathrm{m}$ and $\omega_\mathrm{m}r_\mathrm{d}^2$. Once $\omega_\mathrm{b}$ and $\omega_\mathrm{cdm}$ (and hence $r_\mathrm{d}$) are constrained by the CMB, this translates into a tight positive correlation between $\Omega_\mathrm{m}$ and $\sum m_\nu\propto(\omega_\mathrm{m} - \omega_\mathrm{cb})$. The strong connection between $\sum m_\nu$ and $\Omega_\mathrm{m}$ is discussed extensively in \cite{Loverde24}.

Finally, we consider the case where $(\theta_*,\omega_\mathrm{b},\omega_\mathrm{cdm})$ are all constrained by the CMB. This amounts to the intersection of the two cases considered before: with greater precision on $\Omega_\mathrm{m}$ due to the additional leverage of a distance measurement at $z_*$, with the degeneracy between $\Omega_\nu$ and $\Omega_\mathrm{cb}$ broken, and with $r_\mathrm{d}$ calibrated, we obtain tight constraints on $\sum m_\nu$ and $\Omega_\mathrm{m}$ that approach the limits obtained from the full combination of DESI and CMB-nl (CMB without lensing), as shown respectively by the purple and black open contours in \cref{fig:sources_bao_cmb}. As expected, the latter combination offers still more constraining power, because CMB-nl data depend on neutrino masses through other effects, such as the integrated Sachs–Wolfe effect, the lensing-induced smoothing of the acoustic peaks, and the damping tail. Among these, the lensing effect is particularly potent.

\begin{figure}
    \centering
    \includegraphics[width=0.95\linewidth]{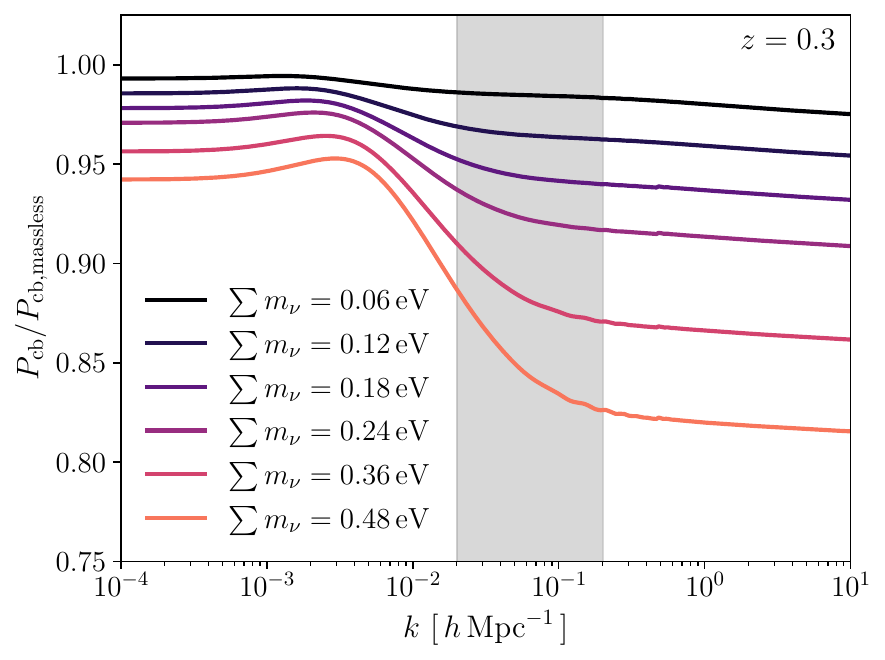}
    \caption{The neutrino effect at $z=0.3$ on the linear power spectrum of cold dark matter and baryons, $P_\mathrm{cb}(k)$, compared to the massless case, for fixed cosmological parameters, $(h,\omega_\mathrm{b},\omega_\mathrm{cdm},A_\mathrm{s},n_\mathrm{s},\tau)$. The range of scales used in the DESI full-shape power spectrum analysis is shown as a gray band.}
    \label{fig:sources_fullshape1}
\end{figure}

\subsection{Growth of structure}\label{sec:free_streaming}

Another prominent effect of massive neutrinos is a suppression of the growth of cosmic structure on small scales. This effect is caused by the thermal phase-space distribution, as neutrinos escape regions smaller than a typical free-streaming length due to their large thermal velocities. Since neutrinos still contribute to the Hubble expansion, as in \cref{eq:late_hubble}, but cluster less efficiently on small scales, the growth of density perturbations is reduced. This produces a characteristic scale-dependent suppression of the matter power spectrum, $P_\mathrm{m}(k)$, as depicted in \cref{fig:sources_fullshape1}. The magnitude of this effect, $\Delta P_\mathrm{m}/P_\mathrm{m}\approx-8f_\nu$ \cite{1998PhRvL..80.5255H,Kiakotou08} (with the precise scaling dependent on which parameters are kept fixed), is significant even for neutrino mass fractions, $f_\nu\equiv \Omega_\nu/\Omega_\mathrm{m}$, well below one percent. 

BAO measurements are blind to neutrino free streaming. However, this information can be captured by properly modeling the full-shape galaxy power spectrum. We will present constraints from the full-shape analysis of DR1 galaxies in \cref{sec:full_shape_sources}.

\section{Standard neutrino results}
\label{Standard}

In this section, we report our baseline constraints on the sum of neutrino masses, $\sum m_\nu$, and on the effective number of relativistic species, $N_\mathrm{eff}$, in the early Universe. We also present results on the preference for the normal mass ordering and constraints on the lightest neutrino mass. Finally, we investigate the impact of different CMB likelihoods and supernova datasets.

\subsection{Baseline neutrino mass constraints}\label{sec:baseline_results}

\begin{figure}
    \centering
    \resizebox{\linewidth}{!}{
        \includegraphics[height=0.8\linewidth]{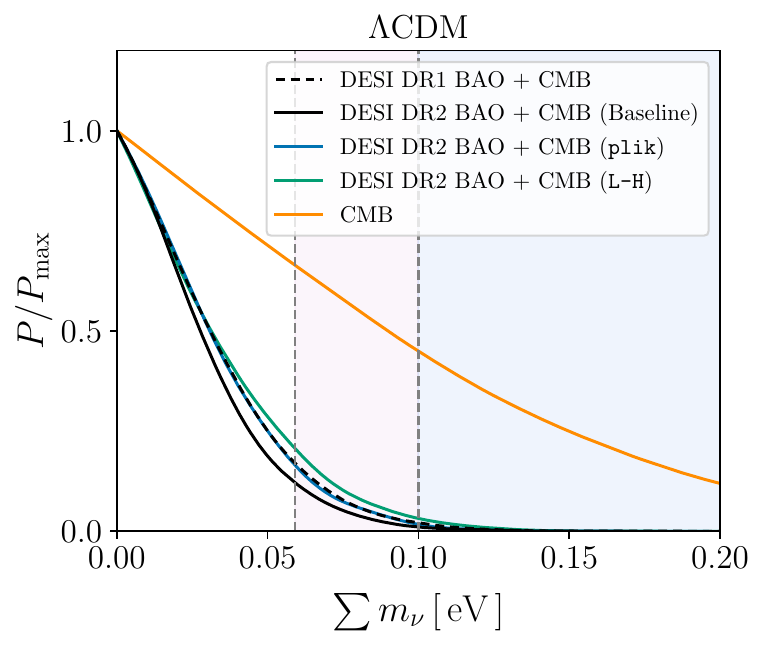}
    }
    \caption{Constraints on $\sum m_\nu$ from different combinations of DESI BAO and CMB data. The CMB dataset includes \emph{Planck} low-$\ell$ TTEE \cite{PlanckLikelihood2020} and high-$\ell$ TTTEEE (PR4 \texttt{CamSpec}) \cite{Rosenberg22,Efstathiou2021} and \emph{Planck} \cite{PlanckLensing2022} and ACT \cite{ACTDR62024} lensing. Shown are results for the CMB alone, CMB combined with DESI DR1 BAO or DR2 BAO, where the latter is our baseline result. Also shown are variations from the baseline for alternative CMB likelihoods: PR3 \texttt{plik} \cite{PlanckLikelihood2020} or PR4 \texttt{L-H} \cite{Tristram2021,Tristram2024}. The vertical dashed lines and shaded regions indicate, from left to right, the minimum masses corresponding to the normal and inverted mass ordering scenarios. 
    }
    \label{fig:baseline_results}
\end{figure}

\begin{figure*}
    \centering
    \includegraphics[height=0.85\linewidth]{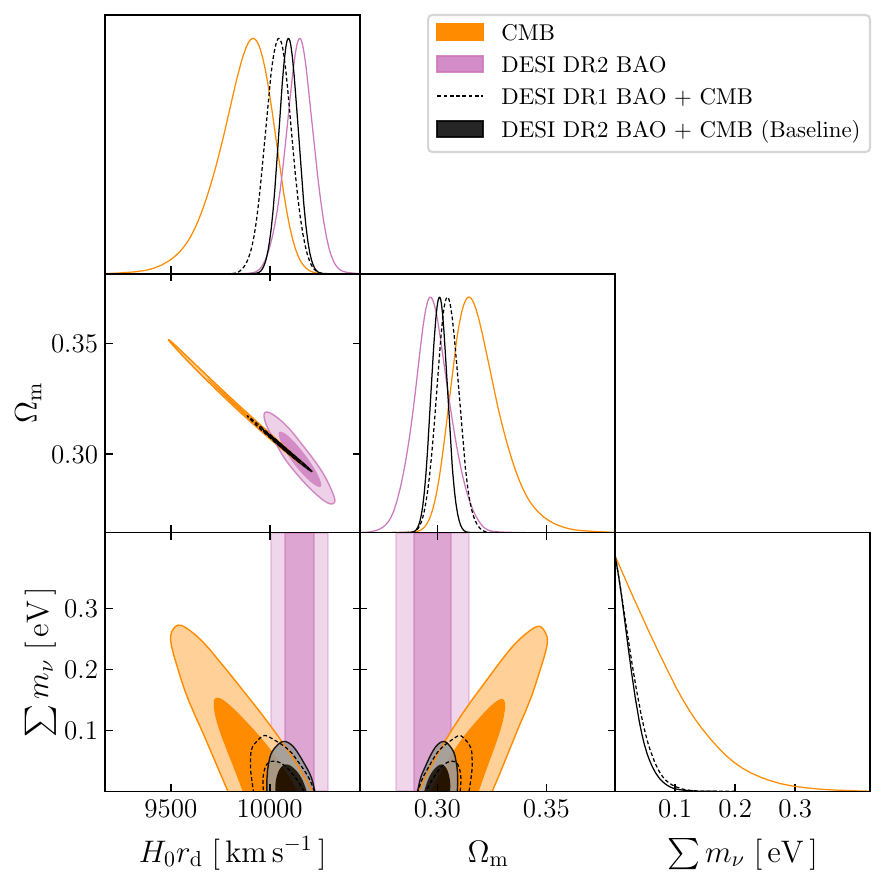}
    \caption{Constraints on the parameter combination $H_0r_\mathrm{d}$ and on the matter density, $\Omega_\mathrm{m}$, both of which are measured to high precision by DESI BAO, and on the sum of neutrino masses, $\sum m_\nu$, from the combination of DESI BAO and CMB data (both for DR1 and the new DR2) and from CMB data alone. The contours represent the 68\% and 95\% probability regions. The figure shows how the DESI preference for lower $\Omega_\mathrm{m}$ and higher $H_0r_\mathrm{d}$, compared to the CMB, leads to upper limits on the sum of neutrino masses that are stronger than expected from forecasts.}
    \label{fig:baseline_results_triangle}
\end{figure*}

We first report the baseline constraint on the sum of neutrino masses, $\sum m_\nu$, described in the companion paper \cite{DESI.DR2.BAO.cosmo}. Assuming the $\Lambda$CDM model and adopting a degenerate mass spectrum accompanied only with a prior that $\sum m_\nu>0$, the combination of DESI DR2 BAO with CMB data yields an upper limit of
\begin{flalign}
\begin{aligned}
    &\qquad \text{DESI DR2 BAO + CMB:} \\
    &\qquad \, \sum m_\nu < \SI{0.0642}{\eV} \quad (95\%), \label{eq:baseline_constraint}
\end{aligned}
&&
\end{flalign}

\noindent
with a marginalized uncertainty of $\sigma(\sum m_{\nu})=\SI{0.020}{\eV}$. As will be discussed below, this is stronger than expected from forecasts \cite{FontRibera14,Brinckmann19}. This constraint is obtained using our baseline CMB dataset, which we refer to simply as `CMB' throughout the paper. This dataset includes both CMB lensing reconstruction from \emph{Planck} PR4 and ACT \cite{ACTDR62024, ACTLensing2024, PlanckLensing2022}, the low-$\ell$ \texttt{SimAll} and \texttt{Commander} likelihoods \cite{PlanckLikelihood2020} and the high-$\ell$ \texttt{CamSpec} likelihood \cite{Efstathiou2021}. The corresponding marginalized posterior distribution is shown in \cref{fig:baseline_results}, demonstrating the $\sim20\%$ improvement compared to DESI DR1 BAO.

The power of CMB and BAO measurements to constrain the sum of neutrino masses is illustrated by \cref{fig:baseline_results_triangle}. Within the $\Lambda$CDM model, there is a geometric degeneracy between CMB constraints on $\sum m_\nu$ and the basic parameters $H_0$ and $\Omega_\mathrm{m}$, as each impact the cosmic expansion rate at late times, thereby changing the distance to the surface of last scattering and shifting the angular size of CMB anisotropies. This degeneracy can be broken with measurements of the BAO distance scale at late times. The figure demonstrates that DESI DR1 BAO measurements prefer values of $H_0r_\mathrm{d}$ and $\Omega_\mathrm{m}$ in the tail of the CMB posteriors, leading to a preference for small masses along the degeneracy direction. This tendency is even stronger for DESI DR2 BAO, giving rise to the upper limit of \cref{eq:baseline_constraint}, which is stronger than was expected a priori.

This limit violates the lower limit from neutrino oscillations for the inverted ordering ($\sum m_\nu\geq\SI{0.10}{\eV}$) and approaches the lower limit for the normal ordering ($\sum m_\nu\geq\SI{0.059}{\eV}$). The consequences of this will be explored in detail in \cref{sec:neutrino_mass_ordering,sec:neutrino_tension}.

However, these results depend on the assumed $\Lambda$CDM model. Generalizing to a dark energy model in which the equation of state, $w$, is constant but may be different from $-1$, we obtain
\begin{flalign}
\begin{aligned}
    & \qquad \text{$w$CDM: DESI DR2 BAO + CMB + DESY5:} \\
    &\qquad \left\{
    \begin{aligned}
    \begin{split}
         \sum m_\nu &< \SI{0.0586}{\eV}  \\
         w &= -0.961^{+0.041}_{-0.043}
    \end{split}\quad (95\%),
    \end{aligned}
    \right.
\end{aligned}
\end{flalign}

\noindent
where we additionally added supernovae from the DESY5 dataset to further constrain the equation of state. We obtain values, $w\approx-1$, consistent with a cosmological constant, and a neutrino mass limit that is even tighter than in $\Lambda$CDM. Similar results are found for supernovae from Union3 and Pantheon+ (see \cref{tab:neutrino_constraints}). The reason that the constraints tighten for $w$CDM, despite adding a parameter, is that the expansion rate increases at late times for $w>-1$ , worsening the tension between DESI BAO and CMB, which can be compensated by smaller values of $\sum m_\nu$. It is only when we allow for a varying dark energy equation of state, parametrized by $w_0$ and $w_a$ (as defined through \cref{eq:w0wa_definition,eq:rho_DE}), that we obtain a relaxed limit
\begin{flalign}
\begin{aligned}
    &  \text{$w_0w_a$CDM: DESI DR2 BAO + CMB + DESY5:} \\
    & \left\{
    \begin{aligned}
    \begin{split}
         &\sum m_\nu < \SI{0.129}{\eV}\\
         &w_0 = -0.76^{+0.12}_{-0.11}\\
         &w_a = -0.82^{+0.46}_{-0.48}
    \end{split}\quad (95\%).
    \end{aligned}
    \right.
\end{aligned}
\end{flalign}

\begin{figure*}
    \centering
    \resizebox{\linewidth}{!}{
        \includegraphics[height=0.4\linewidth]{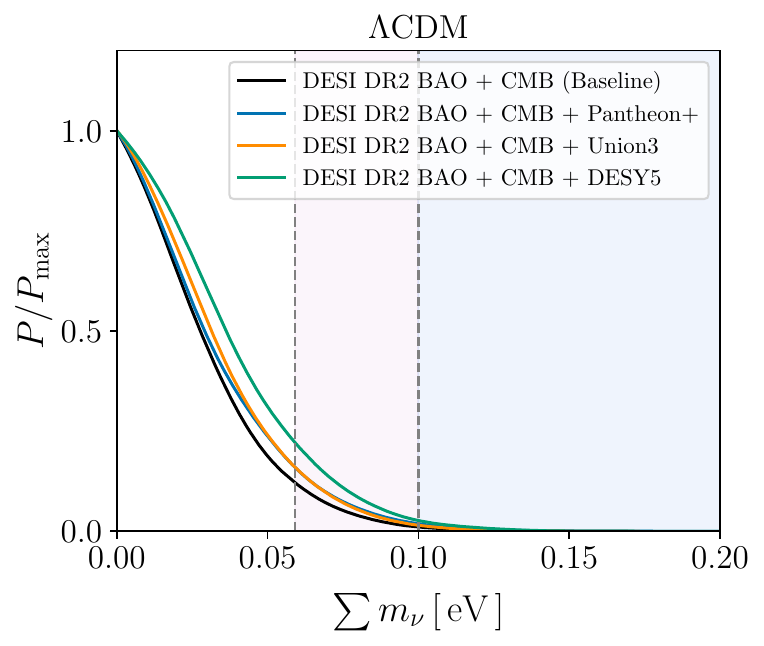}
        \includegraphics[height=0.4\linewidth]{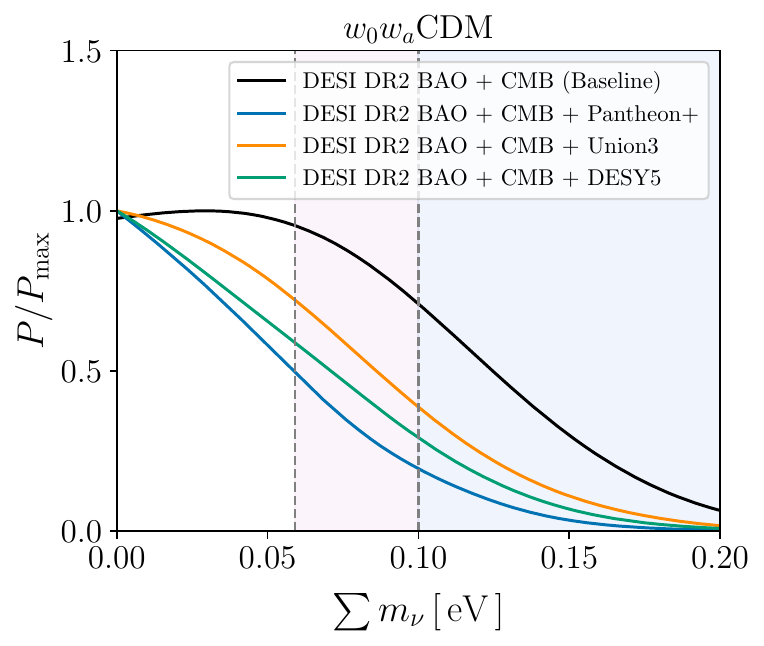}
    }
    \caption{Constraints on $\sum m_\nu$ from our baseline combination (DESI DR2 BAO with CMB) and with the addition of different SNe datasets, Pantheon+ \citep{PantheonPlus2022,Brout:2022}, Union3 \citep{Union32023} and DESY5 \citep{DESY5SN2024}, for $\Lambda$CDM (left) and $w_0w_a$CDM (right). The impact of the choice of SNe sample is significantly greater in $w_0w_a$CDM, while the choice of CMB likelihood is more important in $\Lambda$CDM.}
    \label{fig:baseline_results_sn}
\end{figure*}

\noindent
This upper limit remains consistent with the lower limit from neutrino oscillation measurements for both mass orderings. Although the marginalized posterior distribution of $\sum m_\nu$, shown in the right panel of \cref{fig:baseline_results_sn}, still peaks at the prior edge, $\sum m_\nu=0$, our analyses based on profile likelihoods and on models with effective neutrino masses in \cref{sec:neutrino_tension}, will show that moving to $w_0w_a$CDM shifts the most likely value of $\sum m_\nu$ in the positive direction. The connection between evolving dark energy and neutrino masses will be discussed further in that section.

\subsection{Effective number of relativistic species}\label{sec:Neff}

DESI BAO and CMB can also constrain the effective number of relativistic species in the early Universe, $N_\mathrm{eff}$, defined through \cref{eq:Neff_def}. For the one-parameter extension $\Lambda$CDM+$N_\mathrm{eff}$, we obtain
\begin{flalign}
\begin{aligned}
    &\qquad \text{DESI DR2 BAO + CMB:} \\
    &\qquad N_\mathrm{eff} = 3.23^{+0.35}_{-0.34}\quad (95\%). 
\end{aligned}
&&
\end{flalign}

\noindent
This value is higher than in the DR1 analysis \cite{DESI2024.VI.KP7A}, which we similarly attribute to the preference for larger $H_0r_\mathrm{d}$ and smaller $\Omega_\mathrm{m}$. In the more general case, where we constrain $\sum m_\nu$ and $N_\mathrm{eff}$ simultaneously, we find
\begin{flalign}
\begin{aligned}
    & \qquad \text{DESI DR2 BAO + CMB:} \\
    &\qquad \left\{
    \begin{aligned}
    \begin{split}
        \sum m_\nu  &< \SI{0.0741}{\eV} \\
        N_\mathrm{eff} &= 3.16^{+0.34}_{-0.33}
    \end{split} \quad (95\%).
    \end{aligned}
    \right.
\end{aligned}
&&
\end{flalign}

\noindent
Due to the small correlation between the two parameters ($\rho=0.18$), the upper limit on $\sum m_\nu$ only increases by $15\%$. In all cases, our constraints on $N_\mathrm{eff}$ remain compatible with the Standard Model prediction that $N_\mathrm{eff}=3.044$, as detailed in \cref{tab:neutrino_constraints}.

In addition to the geometrical signature of $N_\mathrm{eff}$ probed by our standard analysis of DESI BAO and CMB data, neutrinos also induce a phase shift in the BAO that has already been detected with DESI DR1 data \cite{Whitford24}. The origin of this effect lies in a temporal shift in the gravitational potential at the time the BAO propagate through space, which occurs because neutrinos carry energy and move faster than the sound speed of the primordial plasma (see \cite{Bashinsky04,Baumann16} for details). It was first measured in \cite{Baumann19}, where the phase shift was expressed as $\phi(k)=\beta_\phi(N_\mathrm{eff})f(k)$, where $f(k)$ is a template that captures the expected scale dependence. In the Standard Model with $N_{\mathrm{eff}} = 3.044$, it is expected that the phase shift amplitude in this parameterization, $\beta_{\phi}$, should be $\beta_{\phi} = 1$; $\beta_{\phi} = 0$ would correspond to no detection of a phase shift and $N_{\mathrm{eff}} = 0$. In \cite{Whitford24}, $\beta_{\phi}$ was measured using DESI DR1 BAO data, giving $\beta_{\phi} > 0$ at $1.6\sigma$ from a combination of tracers. Repeating this analysis with DESI DR2 BAO gives $\beta_{\phi} > 0$ at $2.6\sigma$. Independently of the CMB, this analysis provides a hint of a positive detection of the phase shift due to neutrinos ($N_\mathrm{eff}>0$), but no meaningful constraint on $N_\mathrm{eff}$ can be placed without combining with external datasets or priors.

\begin{figure}
    \centering
    \resizebox{\linewidth}{!}{
        \includegraphics[width=0.9\linewidth]{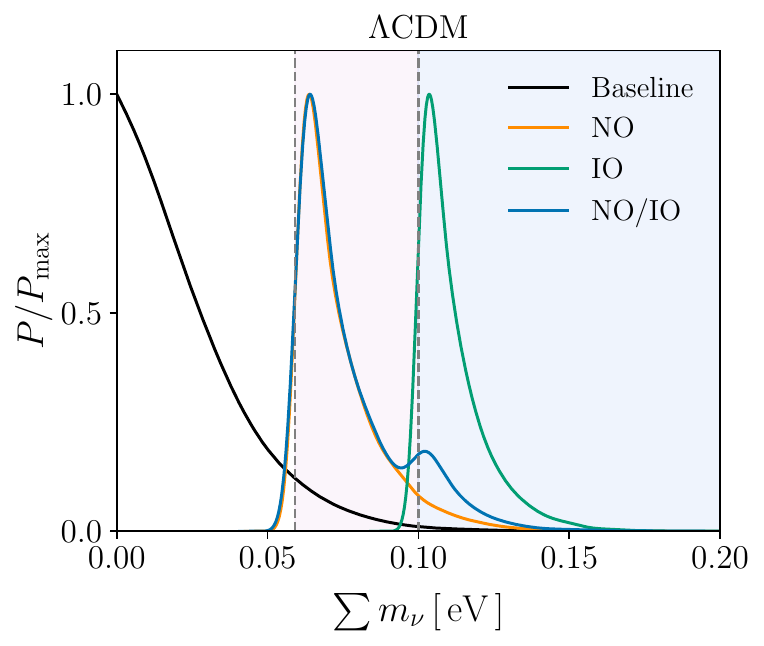}
    }
    \caption{Constraints on the sum of neutrino masses, $\sum m_\nu$, for different assumptions about the neutrino mass ordering using the combination of DESI BAO, CMB, and a global fit to neutrino oscillation experiments (NuFIT 6.0) \cite{Esteban:2024eli}, except in the baseline case which excludes NuFIT. The offset between the curves and the vertical dashed lines corresponding to the oscillation lower limits is an artefact of smoothing.}
    \label{fig:oscillation_results}
\end{figure}

\subsection{Neutrino mass ordering}\label{sec:neutrino_mass_ordering}

Our baseline analysis assumes that $\sum m_\nu = 3m_\nu$, where $m_\nu$ is the mass of three degenerate neutrino mass eigenstates. While this is a reasonable approximation for the cosmological effects of massive neutrinos \cite{Vagnozzi17,Archidiacono20,Lesgourgues05,DiValentino18,Herold24}, it is clearly incompatible with the non-zero squared mass differences implied by the finding of neutrino oscillations. An alternative approach is to adopt a parametrization for the sum of neutrino masses in terms of the lightest neutrino mass, $m_l$, and the squared mass splittings $\Delta m^2_{21}$ and $\rvert\Delta m^2_{31}\rvert$ \cite{Loureiro19, Choudhury20}. In the case of the normal ordering (NO), we have
\begin{align}
    \textstyle\sum m_\nu\!=m_1 + \sqrt{m_1^2 + \Delta m^2_{21}} + \sqrt{m_1^2+\Delta m^2_{31}},
\end{align}

\noindent
with $m_l = m_1<m_2<m_3$, while in the case of the inverted ordering (IO),
\begin{align}
    \textstyle\sum m_\nu\!=m_3 &+ \sqrt{m_3^2-\Delta m^2_{31}} + \sqrt{m_3^2-\Delta m^2_{32}},
\end{align}

\noindent
with $m_l = m_3<m_1<m_2$. To be completely general, we may also introduce a binary variable, $\mathcal{M}$, for the mass ordering. In this most general case with $\mathcal{M}$ free (denoted as NO/IO), we will assign equal prior probability to both mass orderings. While this prescription offers a natural way to fold in prior information from neutrino oscillation experiments, it also leads to highly non-linear priors for the heavier neutrino masses. If the same mechanism is responsible for generating all three neutrino masses, then a prior that is linear for all masses may be better motivated.

From the combination of DESI BAO, CMB, and a global fit to neutrino oscillation experiments (NuFIT 6.0) \cite{Esteban:2024eli}, we obtain the following constraint on the lightest neutrino mass,
\begin{align}
    m_l < \SI{0.023}{\eV} \;\;\; (95\%;\;\; \text{NO or NO/IO}),
\end{align}

\noindent
when assuming the normal mass ordering or in the general case (NO/IO). When assuming the inverted mass ordering, we find a very similar limit of
\begin{align}
    m_l < \SI{0.024}{\eV} \;\;\; (95\%;\;\; \text{IO}).
\end{align}

\noindent
This is a significant improvement with respect to a similar analysis utilizing BOSS DR12 \cite{2016MNRAS.455.1553R}, \emph{Planck} 2015 \cite{2016A&A...594A...1P}, Pantheon SNe Ia \cite{2018ApJ...859..101S}, and BBN information \cite{2018PhR...754....1P}, that yielded $m_l < \SI{0.086}{\eV}$ (95\%) \cite{Loureiro19}. The result may also be compared with the constraint, $m_l<\SI{0.040}{\eV}$ \cite{Choudhury20} from \emph{Planck} 2018 \cite{PlanckCosmology2020}, BOSS DR12 \cite{2016MNRAS.455.1553R}, the DR7 Main Galaxy Survey \cite{Ross15}, and the Six-degree-Field Galaxy Survey (6dFGS) \cite{Beutler12}.

In the general case, the data prefer the normal mass ordering. Assuming $\Lambda$CDM, we find a posterior probability from DESI BAO + CMB + NuFIT of
\begin{align}
    P(\text{NO}) = 1 - P(\text{IO}) = 0.91.
\end{align}

\noindent
This corresponds to a Bayes factor of $K=10$. The evidence is slightly weaker for the alternative CMB likelihoods ($K=6$ for \texttt{L-H} and $K=8$ for \texttt{plik}). Overall, this analysis thus provides substantial evidence (according to the Jeffreys scale) in support of the normal mass ordering, under the assumption of the $\Lambda$CDM + $\sum m_\nu$ cosmology. See \cref{fig:oscillation_results} for the marginalized posterior distributions on the sum of neutrino masses for the different mass ordering scenarios.

In a previous DESI analysis based on DR1 BAO data \cite{DESI2024.VI.KP7A}, the upper limits for the normal and inverted mass orderings were determined by assuming a degenerate mass spectrum (as in the baseline case here) and imposing the additional prior that $\sumnu \geq\SI{0.059}{\eV}$ (NO) or $\sumnu \geq\SI{0.10}{\eV}$ (IO). The posteriors obtained under this approximation agree well in the tail of the distribution. Consequently, we confirm that the approximate procedure produces accurate $95\%$ upper limits. In the case of the normal ordering, we find
\begin{align}
    \sum m_\nu &< 0.101\,\si{\eV} \;\;\; (95\%;\;\; \text{NO}),\\
    \sum m_\nu &< 0.105\,\si{\eV} \;\;\; (95\%;\;\; \sum m_\nu\geq\SI{0.059}{\eV}),
\end{align}

\noindent
while in the case of the inverted mass ordering
\begin{align}
    \sum m_\nu &< 0.133\,\si{\eV} \;\;\; (95\%;\;\; \text{IO}),\\
    \sum m_\nu &< 0.135\,\si{\eV} \;\;\; (95\%;\;\; \sum m_\nu\geq\SI{0.10}{\eV}),
\end{align}

\noindent
thus validating the results from \cite{DESI2024.VI.KP7A}.

\begin{table*}
\centering
\resizebox{\linewidth}{!}{
    \begin{tabular}{lcccccc}
    \toprule
    Model/Dataset & $\Omega_\mathrm{m}$ & $H_0$ [km s$^{-1}$ Mpc$^{-1}$] & $\sum m_\nu$ [eV] & $N_\mathrm{eff}$ & $w$ or $w_0$ & $w_a$ \\
    \midrule
    $\mathbf{\Lambda}$\textbf{CDM+}$\mathbf{\sum m_\nu}$ &  &  &  &  &  &  \\
    DESI BAO+CMB (Baseline) & $0.3009\pm 0.0037$ & $68.36\pm 0.29$ & $<0.0642$ & --- & --- & --- \\
    DESI BAO+CMB (\texttt{L-H}) & $0.2995\pm 0.0037$ & $68.48\pm 0.30$ & $<0.0774$ & --- & --- & --- \\
    DESI BAO+CMB (\texttt{plik}) & $0.2998\pm 0.0038$ & $68.56\pm 0.31$ & $<0.0691$ & --- & --- & --- \\
    DESI BAO+CMB+Pantheon+ & $0.3021\pm 0.0036$ & $68.27\pm 0.29$ & $<0.0704$ & --- & --- & --- \\
    DESI BAO+CMB+Union3 & $0.3020\pm 0.0037$ & $68.28\pm 0.29$ & $<0.0674$ & --- & --- & --- \\
    DESI BAO+CMB+DESY5 & $0.3036\pm 0.0037$ & $68.16\pm 0.29$ & $<0.0744$ & --- & --- & --- \\
    \hline
    $\mathbf{\Lambda}$\textbf{CDM+}$\mathbf{N_\mathrm{eff}}$ &  &  &  &  &  &  \\
    DESI BAO+CMB & $0.3004\pm 0.0042$ & $69.2\pm 1.0$ & --- & $3.23\pm 0.18$ & --- & --- \\
    \hline
    $\mathbf{\Lambda}$\textbf{CDM+}$\mathbf{\sum m_\nu}$\textbf{+}$\mathbf{N_\mathrm{eff}}$ &  &  &  &  &  &  \\
    DESI BAO+CMB & $0.2996\pm 0.0042$ & $69.00\pm 0.97$ & $<0.0741$ & $3.16\pm 0.17$ & --- & --- \\
    \hline
    $\mathbf{w}$\textbf{CDM+}$\mathbf{\sum m_\nu}$ &  &  &  &  &  &  \\
    DESI BAO+CMB & $0.2943\pm 0.0073$ & $69.28\pm 0.92$ & $<0.0851$ & --- & $-1.039\pm 0.037$ & --- \\
    DESI BAO+CMB+Pantheon+ & $0.3045\pm 0.0051$ & $67.94\pm 0.58$ & $<0.0653$ & --- & $-0.985\pm 0.023$ & --- \\
    DESI BAO+CMB+Union3 & $0.3047\pm 0.0059$ & $67.93\pm 0.69$ & $<0.0649$ & --- & $-0.985\pm 0.028$ & --- \\
    DESI BAO+CMB+DESY5 & $0.3094\pm 0.0049$ & $67.34\pm 0.53$ & $<0.0586$ & --- & $-0.961\pm 0.021$ & --- \\
    \hline
    $\mathbf{w}$\textbf{CDM+}$\mathbf{N_\mathrm{eff}}$ &  &  &  &  &  &  \\
    DESI BAO+CMB & $0.2932\pm 0.0075$ & $69.8\pm 1.1$ & --- & $3.13\pm 0.19$ & $-1.047\pm 0.040$ & --- \\
    DESI BAO+CMB+Pantheon+ & $0.3039\pm 0.0052$ & $68.7\pm 1.0$ & --- & $3.22\pm 0.19$ & $-0.987\pm 0.024$ & --- \\
    DESI BAO+CMB+Union3 & $0.3039\pm 0.0059$ & $68.8\pm 1.0$ & --- & $3.23\pm 0.19$ & $-0.986\pm 0.029$ & --- \\
    DESI BAO+CMB+DESY5 & $0.3087\pm 0.0050$ & $68.35\pm 0.98$ & --- & $3.27\pm 0.18$ & $-0.960\pm 0.022$ & --- \\
    \hline
    $\mathbf{w_0w_a}$\textbf{CDM+}$\mathbf{\sum m_\nu}$ &  &  &  &  &  &  \\
    DESI BAO+CMB & $0.353\pm 0.022$ & $63.7^{+1.7}_{-2.2}$ & $<0.163$ & --- & $-0.42^{+0.24}_{-0.21}$ & $-1.75\pm 0.63$ \\
    DESI BAO+CMB+Pantheon+ & $0.3109\pm 0.0057$ & $67.54\pm 0.59$ & $<0.117$ & --- & $-0.845\pm 0.055$ & $-0.57^{+0.23}_{-0.19}$ \\
    DESI BAO+CMB+Union3 & $0.3269\pm 0.0088$ & $65.96\pm 0.84$ & $<0.139$ & --- & $-0.674\pm 0.090$ & $-1.06^{+0.34}_{-0.28}$ \\
    DESI BAO+CMB+DESY5 & $0.3188\pm 0.0058$ & $66.75\pm 0.56$ & $<0.129$ & --- & $-0.758\pm 0.058$ & $-0.82^{+0.26}_{-0.21}$ \\
    \hline
    $\mathbf{w_0w_a}$\textbf{CDM+}$\mathbf{N_\mathrm{eff}}$ &  &  &  &  &  &  \\
    DESI BAO+CMB & $0.355^{+0.023}_{-0.020}$ & $63.1^{+1.8}_{-2.5}$ & --- & $2.96\pm 0.18$ & $-0.40^{+0.23}_{-0.20}$ & $-1.82\pm 0.60$ \\
    DESI BAO+CMB+Pantheon+ & $0.3113\pm 0.0058$ & $67.5\pm 1.0$ & --- & $3.04\pm 0.19$ & $-0.839\pm 0.056$ & $-0.61^{+0.23}_{-0.20}$ \\
    DESI BAO+CMB+Union3 & $0.3282\pm 0.0088$ & $65.7\pm 1.2$ & --- & $3.00\pm 0.18$ & $-0.663\pm 0.089$ & $-1.11^{+0.32}_{-0.28}$ \\
    DESI BAO+CMB+DESY5 & $0.3193\pm 0.0058$ & $66.6\pm 1.0$ & --- & $3.01\pm 0.19$ & $-0.750\pm 0.058$ & $-0.88^{+0.25}_{-0.22}$ \\
    \bottomrule
    \end{tabular}
}
\caption{Constraints from DESI DR2 BAO on cosmological parameters for models that include the sum of neutrino masses, $\sum m_\nu$, or the effective number of relativistic degrees of freedom, $N_\mathrm{eff}$, as additional free parameters. When $\sum m_\nu$ is varied, we use the $\sum m_\nu>0$ prior. Otherwise, a fixed value of $\sum m_\nu=\SI{0.06}{\eV}$ is assumed. When $N_\mathrm{eff}$ is not varied, we use a fixed value of $N_\mathrm{eff}=3.044$. We report $68\%$ limits for all parameters, except $\sum m_\nu$ for which $95\%$ upper limits are given. The baseline CMB dataset includes the low-$\ell$ \texttt{SimAll} and \texttt{Commander} likelihoods \cite{PlanckLikelihood2020} and the high-$\ell$ \texttt{CamSpec} likelihood \cite{Efstathiou2021} for \emph{Planck} and CMB lensing reconstruction from \emph{Planck} PR4 and ACT \cite{ACTDR62024, ACTLensing2024, PlanckLensing2022}.}

\label{tab:neutrino_constraints}
\end{table*}

\subsection{Impact of CMB likelihoods}\label{sec:cmb_likelihoods}

We investigate the dependence of neutrino mass constraints on the \emph{Planck} CMB likelihood, specifically comparing the \texttt{plik}, \texttt{CamSpec} and \texttt{L-H} combinations within $\Lambda$CDM. The three likelihoods produce some notable differences in their constraints on the broader $\Lambda$CDM parameter space, the most striking being a preference for a smaller $\omega_{\mathrm b}$ in \texttt{CamSpec} and \texttt{L-H} in comparison to \texttt{plik}. In addition, the \texttt{L-H} likelihoods produce smaller degeneracies between the primordial power spectrum amplitude, $A_{\mathrm s}$, and optical depth, $\tau$. These differences, while important in their own respect, do not appear to be directly responsible for the offsets between the resulting constraints on the summed neutrino mass.

One of the key differences in the three likelihoods, that may be driving the subtle differences in neutrino mass, are their reported measurements of $A_{\mathrm{lens}}$ \citep{Calabrese2008,Renzi18,Mokeddem23}, a phenomenological parameter that scales the amplitude of the lensing potential. Original results from \texttt{plik} resulted in values of $A_\mathrm{lens}>1$ at greater than $2\sigma$ significance. Subsequent analysis with \texttt{CamSpec} reduced this anomaly. For \texttt{L-H}, the obtained values of $A_{\mathrm{lens}}$ are slightly smaller still and consistent with unity. The exact causes of the lensing anomaly are complex and may be due to a combination of choices in data processing and modeling.

In combination with DESI DR2 BAO, the \texttt{plik} and \texttt{CamSpec} likelihoods from \emph{Planck} provide similar constraints of $\sum m_{\nu}<\SI{0.0691}{\eV}$ and $\sum m_{\nu}<\SI{0.0642}{\eV}$, respectively. Although $A_\mathrm{lens}$ is less discrepant from unity in \texttt{CamSpec} than in \texttt{plik}, perhaps suggesting a relaxation of the neutrino mass limit, \texttt{CamSpec} also uses more data from \emph{Planck} PR4 and provides tighter parameter constraints overall. For \texttt{L-H}, the upper limit is significantly larger, $\sum m_{\nu}<\SI{0.0774}{\eV}$. The broader constraints from \texttt{L-H} remain even when incorporating the squared mass splittings for the normal or inverted mass orderings. See \cref{fig:baseline_results} for a comparison of the 1D marginalized posterior distributions when adopting the different likelihoods.

\subsection{Impact of SNe data}

Within the $\Lambda$CDM framework, we find that the influence of the three ensembles of Type Ia supernovae on neutrino mass is significantly weaker than that of the selected CMB likelihood, provided that BAO data are included. In all cases, the constraints are slightly relaxed compared to the case without SNe, and accompanied by higher values of $\Omega_\mathrm{m}$ and lower values of $H_0$. For our baseline CMB likelihood (\texttt{CamSpec}), Union3 gives the strongest limit of $\sum m_{\nu} < \SI{0.0674}{\eV}$, while Pantheon+ gives $\sumnu < \SI{0.0704}{\eV}$. DESY5 generally pushes the constraints to larger values, yielding $\sum m_{\nu} < \SI{0.0744}{\eV}$. However, within \(\Lambda\)CDM, the offsets between CMB likelihoods dominate (see \cref{tab:neutrino_constraints}). In the absence of DESI BAO, the impact of SNe is more significant \citep[e.g.][]{Loverde24}, but our focus here is on the case with BAO included.

Once we move to a dynamical dark energy in the $w_0$--$w_a$ parameter space, the differences between the CMB likelihoods are relatively smaller. For DESY5, we find similar constraints of $\sum m_{\nu} < \SI{0.129}{\eV}$ for \texttt{plik} and $\sum m_{\nu} < \SI{0.133}{\eV}$ for \texttt{CamSpec}, while for \texttt{L-H} the upper limit is $\sum m_{\nu} < \SI{0.148}{\eV}$. On the other hand, changing the SNe dataset produces considerably larger differences in the constraints, as can be seen from \cref{fig:baseline_results_sn}. With Pantheon+, the upper limit is given by $\sum m_{\nu} < \SI{0.117}{\eV}$, while this is $\sum m_{\nu} < \SI{0.139}{\eV}$ with Union3 and $\sum m_{\nu} < \SI{0.129}{\eV}$ for DESY5. These differences are driven by their relative departures from $\Lambda$CDM, which is largest for Union3 and smallest for Pantheon+ (see \cref{fig:w0wa_mnu_degeneracy} for the case with effective neutrino masses) and their relative constraining power, which is similarly highest for Pantheon+ and lowest for Union3. The departures from $\Lambda$CDM can be related to shifts in $\Omega_\mathrm{m}$ compared to the values preferred by BAO and CMB, when the respective dataset is analyzed in the $\Lambda$CDM model \cite{Tang25}. All three SNe samples are consistent with higher values of $\Omega_\mathrm{m}$, but they are not entirely independent. The high-$z$ sample ($z>0.1$) of DESY5 is largely independent of the other datasets, but its low-$z$ sample ($z<0.1$) is derived from legacy catalogs used by the other compilations. DESY5 also uses a similar methodology as Pantheon+, though with various differences that affect the preference for higher $\Omega_\mathrm{m}$ \cite{DESY5SN2024,Vincenzi25}. Pantheon+ and Union3 have a large fraction of SNe in common, but are methodologically more distinct.

In the case of $w$CDM, Pantheon+ and Union3 present similar upper limits of $\sum m_{\nu} < \SI{0.0653}{\eV}$ and $\sum m_{\nu} < \SI{0.0649}{\eV}$, respectively, while for DESY5, we obtain tighter constraints of $\sum m_{\nu} < \SI{0.0586}{\eV}$.

\section{Neutrino mass tension}\label{sec:neutrino_tension}

\noindent
As discussed in \cref{sec:baseline_results}, our baseline constraint on $\sumnu$ in $\Lambda$CDM approaches but remains compatible with the lower limit from neutrino oscillations under the normal mass ordering. We can thus combine our cosmological results with the constraints from neutrino oscillation experiments. This combination was used in \cref{sec:neutrino_mass_ordering} to constrain the lightest neutrino mass and determine the preference of current data for the normal mass ordering.

However, an unsatisfactory feature of the posterior distribution in our baseline analysis is that it peaks at the edge of the prior, $\sumnu=\SI{0}{\eV}$, and that most of the posterior volume is excluded by neutrino oscillations. Below, we confirm from a frequentist perspective that the data appear to prefer values that lie in the negative mass range. This motivates the analysis in the following subsection, where we use an effective neutrino mass parameter, $\sumnueff$, to explore such scenarios in detail.

\subsection{Profile likelihood}\label{sec:profile_likelihood}

We build profile likelihoods for the sum of neutrino masses by fixing $\sum m_\nu$ to certain values, and then performing maximization of the likelihood over all nuisance parameters and other cosmological parameters.
This frequentist method does not require the use of priors.
In particular, we limit ourselves to probing the physical region where $\sum m_\nu > 0$.
In the case of nuisance parameters in the CMB and the DESI FS likelihoods, we apply penalties as prescribed by their respective frameworks.

In practice, we perform a numerical minimization over the log-likelihood, \textit{i.e.} $-2 \log(\mathcal{L})$, using the \texttt{Minuit}~\cite{JamesMinuit1975} minimizer through its Python frontend, \texttt{iminuit}~\cite{iminuit}.
The number of free parameters ranges from 14 for the DESI BAO + CMB combination to 44 for DESI (FS+BAO) + CMB.

\begin{figure}
  \centering
  \includegraphics[width=0.95\linewidth]{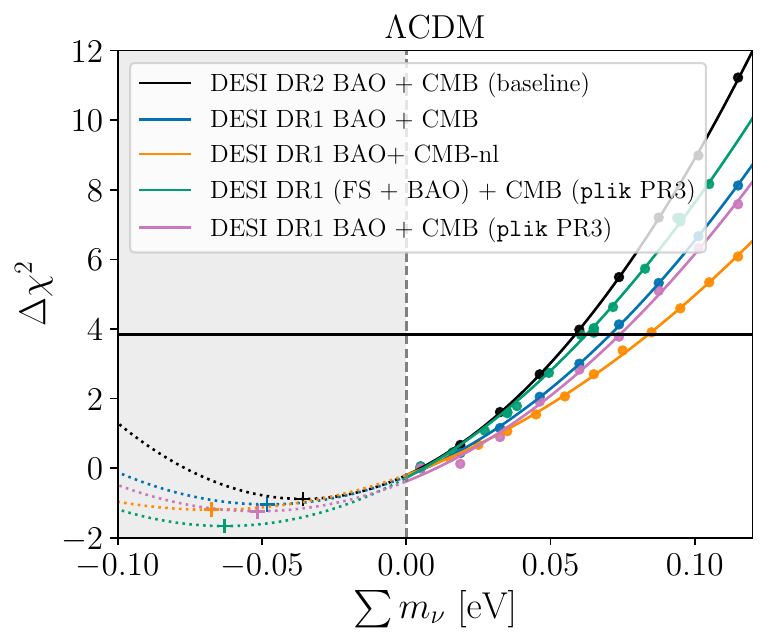}
  \caption{
      Neutrino mass profile likelihood using DESI BAO, BAO+FS, and CMB data with and without the CMB lensing information from \emph{Planck} and ACT DR6. The horizontal line indicates $\Delta \chi^2 = 3.84$, which corresponds to a $p$-value of 0.05.
      In all cases, the apparent minimum of the profile is $\sum m_\nu = 0$, while the fitted parabola shows a minimum in the negatives.
      The value of $\mu_0$ agrees with $\sum m_\nu = 0$ at less than $1\sigma$ for all BAO profiles, and less than $1.5\sigma$ for the full-shape profile.
      When combined with DESI DR1 BAO, CMB PR3 (\texttt{plik}) and PR4 (\texttt{CamSpec}), we obtain extremely similar results.
  }
  \label{fig:lklprofile-BAO-lensing}
\end{figure}

Profile likelihoods for the neutrino mass are often truncated by the $\sum m_\nu > 0$ limit, sometimes not even exhibiting a trough in the positive sector.
Instead, the apparent minimum is located at the physical boundary, $\sum m_\nu = 0$. 
Since we find the resulting profiles to be in good agreement with a parabolic fit, owing to the Gaussian nature of the data (see also \cite{Herold24b,Naredo-Tuero24}), we extend them into a parabola.
The resulting parameters are the parameters of a corresponding $\chi^2$ distribution. 
We report the minimum of the parabola, $\mu_0$, and its scale, $\sigma$, which in turn can be directly interpreted as the constraining power of the likelihood combination.

Considering the overlap between the parabola and the negative sector, we follow the prescription detailed by Feldman and Cousins in~\cite{feldmanUnifiedApproachClassical1998} to build a 95\% confidence level, $\mu_{95}$, based on $\mu_0$ and $\sigma$.
These three parameters are reported for different data combinations in \cref{tab:lklprofile-summary}.

Graphically, $\chi^2\left(\sum m_\nu \right) - \chi^2\left(\sum m_\nu =0\right) = 3.84$ constitutes a good proxy for the Feldman-Cousins upper limit and can be used to visually compare profiles.
Unlike the Feldman-Cousins upper limit, it is not sensitive to the determination of $\mu_0$ by the parabolic fit.
In the figures, we plot $\Delta\chi^2 = \chi^2\left(\sum m_\nu \right) - \chi^2\left(\sum m_\nu^{\rm best} \right)$ as a function of $\sumnu$, where $\sum m_\nu^{\rm best} \geq 0$ is the physically allowed neutrino mass that minimizes the log-likelihood.
Since most of the profiles presented below exhibit a negative $\mu_0$, $\sum m_\nu^{\rm best} = 0$ and the former equation reduces to $\Delta\chi^2 = 3.84$.
Additionally, when the `best-fitting' value of $\sum m_\nu$ is 0, the difference $\chi^2\left(\sum m_\nu = 0 \right) - \chi^2\left(\sum m_\nu = \mu_0 \right)$ informs us about the tension between the recovered minimum and a non-negative neutrino mass.

\begin{figure}
  \centering
  \includegraphics[width=0.95\linewidth]{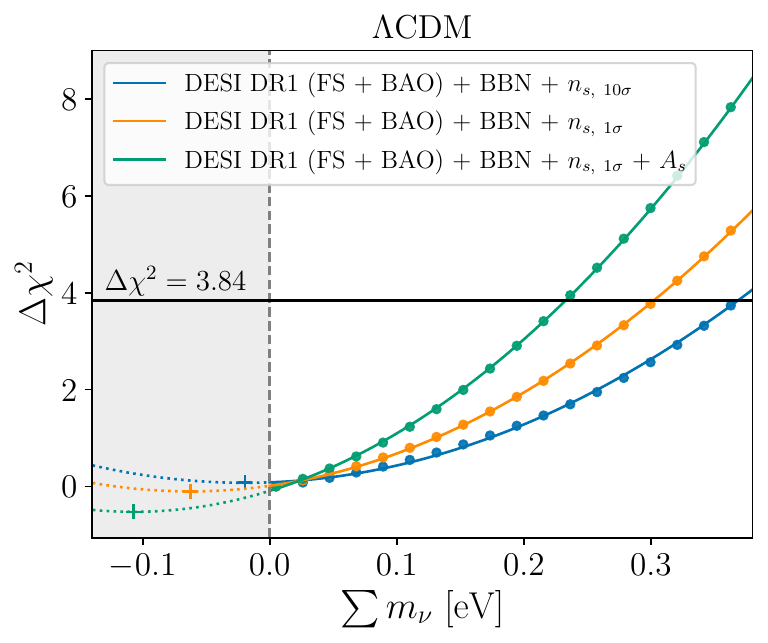}
  \caption{Neutrino mass profile likelihood using the DESI DR1 FS power spectrum analysis, BBN information on $\Omega_\mathrm{b} h^2$, and \emph{Planck} information on $n_\mathrm{s}$. In green, a constraint on $A_\mathrm{s}$ from \emph{Planck} is added. Despite limiting the amount of information from \emph{Planck} in order to avoid potential disagreements between datasets, all minima are still in the negative sector. It can be noted that injecting more information from \emph{Planck} systematically pushes the minimum to more negative values; however, the tension with $\sum m_\nu = 0$ remains below $1\sigma$.
  }
  \label{fig:lklprofile-bbn}
\end{figure}

We first consider combinations of DESI DR1 and DR2 with CMB information.
Profiles are shown in~\cref{fig:lklprofile-BAO-lensing}, and all parameter values are reported in~\cref{tab:lklprofile-summary}.
As could be anticipated from the shape of the Bayesian posteriors in the main analysis, we find that all profiles show a trough firmly in the negative sector once extrapolated to a full parabola.
Using the baseline CMB dataset along with DESI DR1 BAO, we obtain an upper limit of \SI{0.063}{\eV} for the sum of neutrino masses.
Moving from DESI DR1 BAO to DR2 BAO, the constraint improves to
\begin{flalign}
\begin{aligned}
    &\qquad \text{$\Lambda$CDM: DESI DR2 BAO + CMB:} \\
    &\qquad \sum m_\nu < \SI{0.053}{\eV} \quad (95\%), \label{eq:fc_lcdm}
\end{aligned}
&&
\end{flalign}

\noindent
showcasing a \SI{0.010}{\eV} reduction compared to the DR1 result. The constraining power, $\sigma$, also improves by \SI{0.011}{\eV}, as expected from the higher precision of DR2 BAO data. However, the curve shifts in the positive direction by \SI{0.012}{\eV}, limiting the improvement on the upper limit.

Returning to the DESI DR1 BAO + CMB result, we compare this to the same case deprived of CMB lensing information, which yields a larger limit of \SI{0.074}{\eV}. In principle, the constraint on $\sum m_\nu$ comes from a geometrical constraint on $\Omega_\mathrm{m}$, as determined by DESI BAO and CMB measurements, and neutrino free streaming through small-scale suppression of the matter power spectrum, measured through CMB lensing.
Adding CMB lensing information as we do, or the full-shape information from DESI DR1 as we will do next, should thus improve the constraint on the sum of neutrino masses.
This is indeed what we observe here, as the limit decreases by \SI{0.011}{\eV} when adding CMB lensing information.
The profile likelihood technique reveals that this is a two-fold effect, as both $\mu_0$ and $\sigma$ contribute to the determination of the upper limit, $\mu_{95}$.
In this case, the constraining power, $\sigma$, actually improves by \SI{0.023}{\eV}, but this improvement is partly reduced by a shift of the parabola toward the positive region by \SI{0.020}{\eV}. A similar effect is seen in the Bayesian case (see the right panel of \cref{fig:lensing_information}).

We see a different situation when comparing FS + BAO with combinations that only include BAO.
For this comparison, we revert to using the $\texttt{plik}$ \emph{Planck} likelihood as was done for the baseline results in~\cite{DESI2024.VII.KP7B}.
Nevertheless, just like in~\cref{sec:cmb_likelihoods} and~\cite{DESI2024.VII.KP7B} in the Bayesian framework, we find that DESI BAO + CMB \texttt{plik} and \texttt{CamSpec} exhibit extremely similar profiles. We expect an improvement on DESI DR1 BAO when considering the full-shape analysis, which unlike BAO is able to measure the small-scale suppression effect on the matter power spectrum.
As in the previous case, we see an improvement of the upper limit by \SI{0.011}{\eV}. 
The statistical power of the data as measured by $\sigma$ is very close for both curves, with a difference of \SI{0.003}{\eV} in favor of the analysis that includes FS.

These findings show that, in terms of statistical strength, the switch from DESI DR1 BAO to DESI DR2 BAO and the inclusion of CMB lensing are most important, while the improved upper limit in the BAO + FS case is mostly due to a shift toward negative values.

\begin{figure}
  \centering
  \includegraphics[width=0.95\linewidth]{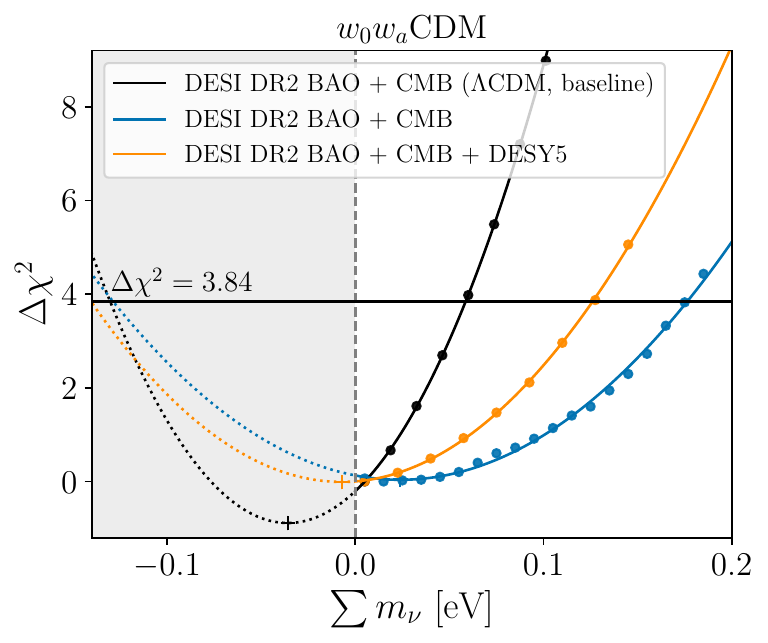}
  \caption{Neutrino mass profile likelihood using DESI DR2 BAO and baseline CMB data, combined with DESY5 supernovae, for $w_0w_a\textrm{CDM}$. 
  When allowing for evolving dark energy, the minimum of the profile shifts very close and even into the positive sector for some data combinations.
  The combination without SNe information can be compared to its $\Lambda\textrm{CDM}$ equivalent: $\sigma$ increases by \SI{0.035}{\eV}, and the upper limit relaxes considerably, aided by the shift toward the positive sector.
  Adding SNe information tightens the parabola again.
  }
  \label{fig:lklprofile-w0waCDM}
\end{figure}

\begin{table*}
  \small
  \centering
  \begin{tabular*}{0.7\textwidth}{@{\extracolsep{\fill}}lccc}
    \toprule
    Model/Dataset &   $\mu_0$ [eV] &   $\sigma$ [eV] &   95\% CL [eV] \\
    \hline
    $\mathbf{\Lambda}$\textbf{CDM+}$\mathbf{\sum m_\nu}$ & & & \\
     DESI DR2 BAO+CMB (\texttt{CamSpec}) & $-0.036$ & $0.043$ & $<0.053$ \\
     DESI DR1 BAO+CMB (\texttt{CamSpec})   &    $-0.048$ &      $0.054$ &     $<0.063$ \\
     DESI DR1 BAO+CMB-nl (\texttt{CamSpec}) &    $-0.068$ &      $0.067$ &     $<0.074$ \\
     DESI DR1 BAO+CMB (\texttt{plik}) & $-0.052$ & $0.056$ & $<0.064$ \\
     DESI DR1 (FS+BAO)+CMB (\texttt{plik}) & $-0.063$ & $0.053$ & $<0.053$ \\
     DESI DR1 (FS+BAO)+BBN+$n_{\mathrm{s},10}$ & $-0.019$ & $0.200$ & $<0.373$ \\
     DESI DR1 (FS+BAO)+BBN+$n_{\mathrm{s},1}$ & $-0.063$ & $0.184$ & $<0.300$ \\
     DESI DR1 (FS+BAO)+BBN+$n_{\mathrm{s},1}$+$A_\mathrm{s}$ & $-0.107$ & $0.163$ & $<0.221$ \\
     \hline
     $\mathbf{w_0w_a}$\textbf{CDM+}$\mathbf{\sum m_\nu}$ & & & \\
     DESI DR2 BAO+CMB & $0.024$ & $0.078$ & $<0.177$ \\
     DESI DR2 BAO+CMB+DESY5 & $-0.007$ & $0.068$ & $<0.126$ \\
     \bottomrule
  \end{tabular*}
  \caption{Profile likelihood parameters for all dataset combinations. Here, $\mu_0$ and $\sigma$ represent the minimum and scale of the parabolic fit, respectively. We also report a 95\% confidence level computed from the Feldman-Cousins prescription. All parameters are given in $\unit{\eV}$.}
  \label{tab:lklprofile-summary}
\end{table*}

DESI can constrain the sum of neutrino masses with limited external information. In \cref{fig:lklprofile-bbn}, we consider the combination of DESI DR1 (FS+BAO) along with a BBN prior on $\Omega_\mathrm{b} h^2$ and CMB information on $n_\mathrm{s}$ and $A_\mathrm{s}$.
The full-shape analysis makes it possible to measure the small-scale suppression effect caused by neutrinos without involving CMB information, while geometric information is provided by the BAO measurement.
If the preference for negative effective neutrino masses is a symptom of some tension between different datasets, then reducing the amount of external information could lessen the preference.

We find that all parabola minima still lie in the negative sector, although the tension with $\sum m_\nu = 0$ is less than $1\sigma$.
Compared to baseline CMB + DESI combinations, the constraints are relaxed, with upper limits as high as \SI{0.373}{\eV} along with $\sigma = \SI{0.200}{\eV}$.
As more stringent CMB information is added to the analysis, such as a narrower constraint on $n_\mathrm{s}$ or additional information on $A_\mathrm{s}$, both constraining power, $\sigma$, and upper limits improve.
The minimum also shifts further toward the negatives by about $\SI{0.045}{\eV}$,
resulting in a decrease in the upper limits of around $\SI{0.07}{\eV}$ for each step.

We now consider the impact of allowing evolving dark energy, under the framework of the $w_0$--$w_a$ parametrization.
DESI data, in combination with external datasets, have been shown to favor dynamical dark energy~\cite{DESI2024.VI.KP7A,DESI2024.VII.KP7B,DESI.DR2.BAO.cosmo,Y3.cpe-s1.Lodha.2025}, especially when including supernova information.
Moving to a dynamical dark energy model helps to alleviate the tension that could be driving the neutrino mass sum toward the negatives.
As shown in~\cref{fig:lklprofile-w0waCDM}, the minima of the parabolas shift very close to the positive sector.
The combination of DESI DR2 BAO and the baseline CMB dataset imposes an upper limit of
\begin{flalign}
\begin{aligned}
    &\qquad \text{$w_0w_a$CDM: DESI DR2 BAO + CMB:} \\
    &\qquad \sum m_\nu < \SI{0.177}{\eV} \quad (95\%), \label{eq:fc_w0wa}
\end{aligned}
&&
\end{flalign}

\noindent
which represents a \SI{0.123}{\eV} relaxation compared to the $\Lambda$CDM case. A large part of this increase is caused by the shift of the parabola toward the positives by \SI{0.06}{\eV}, although further relaxation is expected from the degradation of $\sigma$ from \SI{0.043}{\eV} to \SI{0.078}{\eV}.
When adding SNe information from DESY5, $\sigma$ and $\mu_{95}$ both improve to \SI{0.068}{\eV} and \SI{0.126}{\eV}, while the central value, $\mu_0$, becomes negative again.

Finally, the profile likelihoods can also be used to probe the neutrino mass ordering. 
The profiles are calculated using a degenerate mass approximation with three neutrinos of equal mass, which is still a reasonable approximation for recent DESI data~\cite{Vagnozzi17,Archidiacono20,Lesgourgues05,DiValentino18,Herold24}. 
We use the NuFIT 6.0 constraints \cite{Esteban:2024eli} on the difference of squared masses, from current oscillation experiments, and we consider a situation in which the lightest neutrino has zero mass.
We can then determine a total sum of neutrino masses both in normal and inverted ordering, and compute a $\Delta\chi^2$ between the two.
For baseline CMB + DESI DR2 BAO in $\Lambda\textrm{CDM}$, this procedure yields $\chi^2(\sum m_\nu^{\mathrm{(IO)}}) - \chi^2(\sum m_\nu^{\mathrm{(NO)}}) = 4.6$ in favor of the normal ordering, which is in very good agreement with the Bayes factor reported in~\cref{sec:neutrino_mass_ordering}.

Almost all profiles presented here, especially for $\Lambda\textrm{CDM}$, exhibit a minimum in the negative mass region. Only when introducing dynamical dark energy do we recover minima closer to or inside the positive sector. Nevertheless, when the minima are in the negatives, they mostly remain in agreement with $\sum m_\nu = 0$ to within $1\sigma$. The presented upper limits \cref{eq:fc_lcdm,eq:fc_w0wa} are \SI{0.01}{\eV} lower for the $\Lambda$CDM case and \SI{0.01}{\eV} higher for $w_0w_a$CDM, than those obtained from the Bayesian analysis. 
Considering that we operate with limited statistics, and that our method to exclude the negative region differs, it is not unexpected that we recover somewhat different values. 

\begin{figure*}
    \centering
    \resizebox{\linewidth}{!}{
        \includegraphics[height=0.4\linewidth]{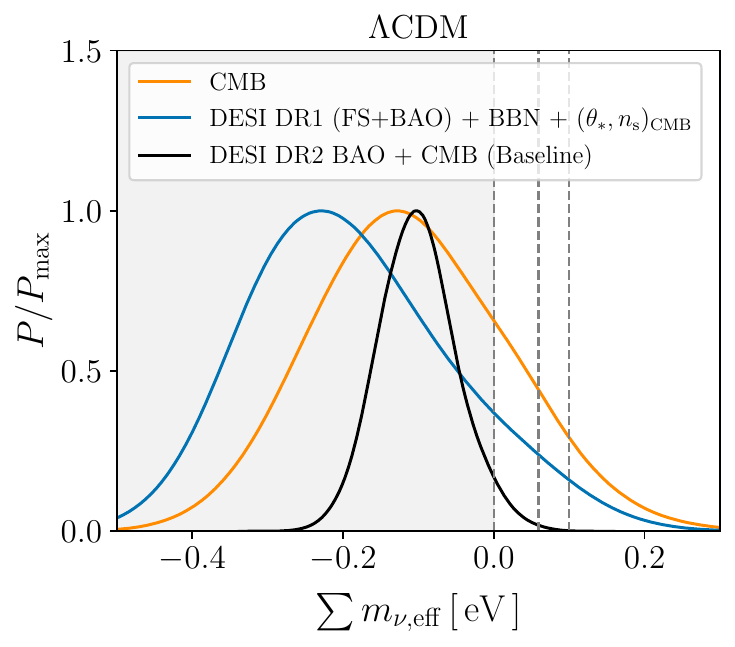}
        \includegraphics[height=0.4\linewidth]{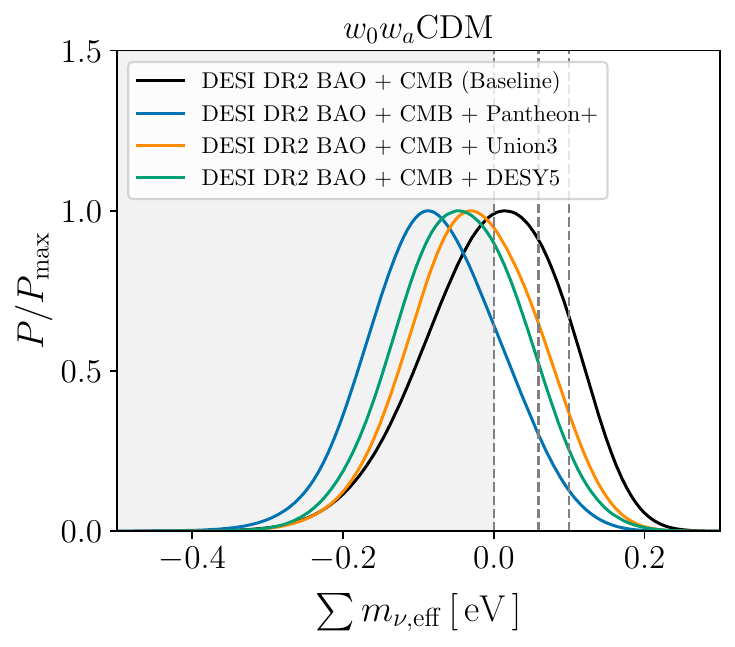}
    }
    \caption{Left: marginalized constraints on the effective neutrino mass parameter, $\sum m_{\nu,\text{eff}}$, in the $\Lambda$CDM model, from the CMB, from a full-shape power spectrum analysis of DESI DR1 (including BBN and CMB priors on $\Omega_\mathrm{b}h^2,\theta_*$, and $n_\mathrm{s}$), and from the combination of DR2 BAO and CMB. All three combinations prefer negative values, but the tension with the lower limits from neutrino oscillations (the second and third vertical dashed lines) is only significant for DESI + CMB. Right: the same for the $w_0w_a$CDM model, using DESI DR2 BAO and CMB data, combined with three different SNe Ia datasets as indicated.}
    \label{fig:negative_mnu_results}
\end{figure*}

\subsection{Effective neutrino masses}\label{sec:effective_neutrinos}

The results of \cref{sec:neutrino_mass_ordering} illustrate that current cosmological constraints on $\sum m_\nu$ are dominated by prior weight effects, at least in the context of the $\Lambda$CDM model. For all cases considered so far, the marginal posterior distribution peaks at the smallest mass allowed by the prior, whether it be $\sum m_\nu=\SI{0}{\eV}$ for a degenerate mass spectrum or $m_l=\SI{0}{\eV}$ when an oscillation-based parametrization is adopted. Furthermore, the profile likelihood analysis shows that our baseline constraints are already in tension with the lower limits from neutrino oscillations and that the minimum of the likelihood is in the negative sector. This motivates a Bayesian analysis with an effective neutrino mass parameter, $\sum m_{\nu,\mathrm{eff}}$, that can be extended to negative values.

We utilize the effective neutrino mass parameter of \cite{Elbers_24}, which allows for a negative neutrino contribution to the energy density at late times. The parameter is implemented at the level of cosmological perturbation theory, ensuring that all cosmological effects of $\sum m_\nu$ are extended consistently to negative values. By definition, it agrees exactly with $\sum m_\nu$ for positive values. We refer to \cref{sec:effective_mass_details} for details. It should be emphasized that this is an \emph{effective} parameter and any evidence for negative values should be interpreted as a signature of unidentified systematic errors or possibly of new physics which may be unrelated to neutrinos, rather than as a direct sign of negative mass neutrinos.

While unorthodox, the effective neutrino mass parameter, $\sum m_{\nu,\mathrm{eff}}$, has a number of advantages compared to the standard approach. First of all, it removes the prior weight effects that drive the posterior away from the maximum likelihood value. It thus provides a robust metric for the tension between cosmological data and oscillation constraints. Moreover, extending the model to negative values provides significant insight, given that the standard analysis reveals only the tail of the distribution of $\sum m_{\nu,\mathrm{eff}}$. This helps, for instance, to differentiate parameter shifts from changes in precision and clarifies the directions of parameter degeneracies.

Adopting the $\sum m_{\nu,\mathrm{eff}}$ parametrization in $\Lambda$CDM, we obtain the following constraint from the combination of DESI DR2 BAO and our baseline CMB dataset:
\begin{flalign}
\begin{aligned}
    &\qquad \text{DESI DR2 BAO + CMB:} \\
    &\qquad \sum m_{\nu,\mathrm{eff}} = -0.101^{+0.047}_{-0.056}\,\si{\eV} \quad (68\%). 
\end{aligned}
&& \label{eq:baseline_negnu}
\end{flalign}

\noindent
The marginalized error is $\sigma(\sum m_{\nu,\mathrm{eff}})=\SI{0.053}{\eV}$. This is a factor $2.6$ larger than in the baseline analysis and more in line with the expectation from forecasts \cite{FontRibera14,Brinckmann19}. \cref{eq:baseline_negnu} amounts to a slight shift and a $20\%$ improvement in precision compared to the constraint, $\sum m_{\nu,\mathrm{eff}}=-0.125_{-0.070}^{+0.058}\,\si{\eV}$, from \cite{Elbers_24} obtained for DESI DR1 BAO and CMB data. The tension with the lower limit, $\sum m_\nu\geq\SI{0.059}{\eV}$, for the normal mass ordering is $3.0\sigma$ (an increase from the $2.8\sigma$ quoted in \cite{Elbers_24}).\footnote{We compute the tension in terms of the probability to exceed the lower limit from neutrino oscillations, using the 1D marginalized posterior distribution of $\sum m_{\nu,\mathrm{eff}}$. See \cref{sec:tension_metrics} for alternative tension metrics.} The marginal posterior distribution is shown in the left panel of \cref{fig:negative_mnu_results} and the constraints on other parameters are given in \cref{tab:neg_neutrino_constraints}. As the asymmetric error bars in \cref{eq:baseline_negnu} indicate, the posterior is non-Gaussian and slightly skewed toward negative values.

\begin{figure*}
  \centering
  \resizebox{\linewidth}{!}{
    \includegraphics[height=0.4\linewidth]{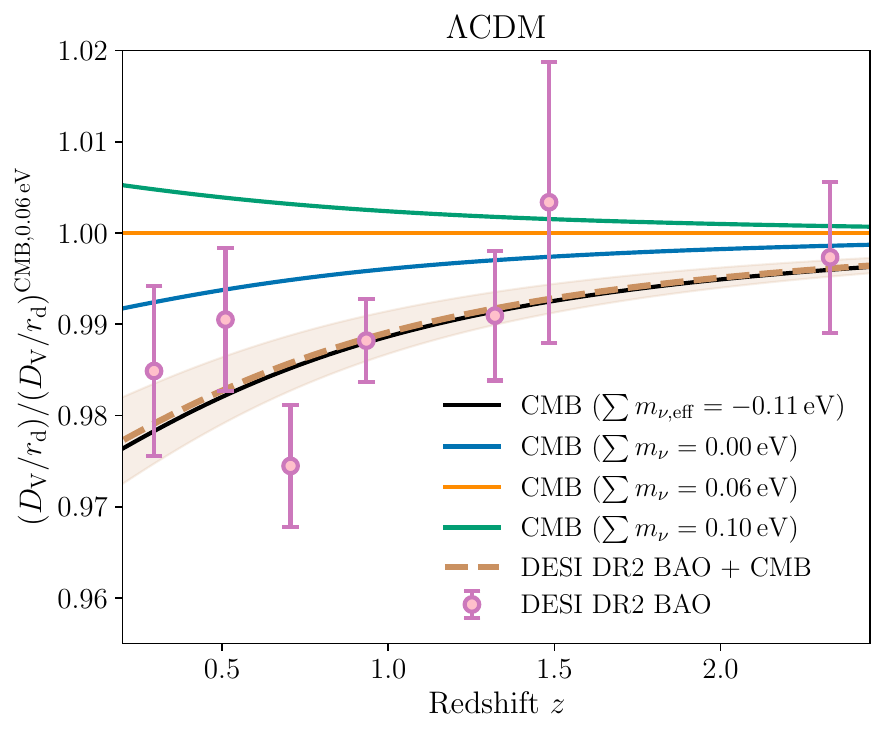}
    \includegraphics[height=0.4\linewidth]{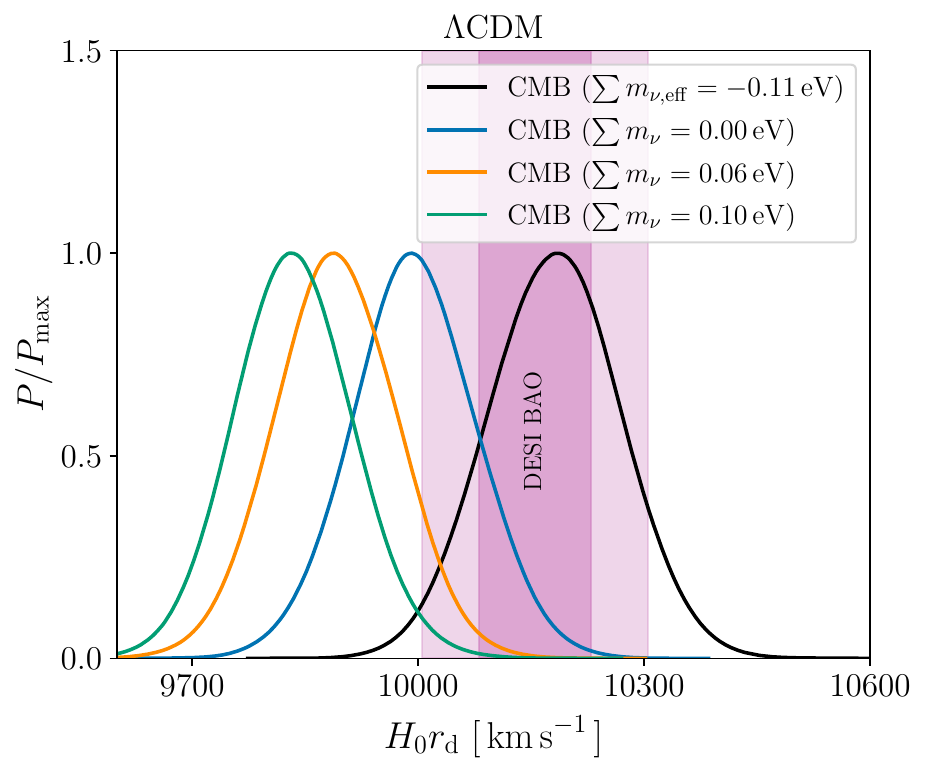}
  }
  \caption{Left: Posterior mean predictions from our baseline CMB dataset for the isotropic BAO distance measurements, $D_\mathrm{V}/r_\mathrm{d}$. We use the $\Lambda$CDM model with fixed values of the effective sum of neutrino masses, $\sum m_{\nu,\mathrm{eff}} = -0.11,\, 0,\, 0.06,$ and $\SI{0.1}{\eV}$. We also show the DESI DR2 BAO data points and the posterior prediction for DESI + CMB with $\sum m_{\nu,\mathrm{eff}}$ free as a dashed line. For visual clarity, we only show the $1\sigma$ uncertainty for DESI + CMB. Right: Marginalized 1D posterior constraints on $H_0 r_\mathrm{d}$ from our baseline CMB dataset for the same fixed values of $\sum m_{\nu,\mathrm{eff}}$.  The vertical shaded regions indicate the 68\% and 95\% constraints from DESI DR2 BAO. The amplitude of $D_\mathrm{V}/r_\mathrm{d}$ is inversely proportional to $H_0r_\mathrm{d}$. These plots show how the CMB preference for smaller $H_0 r_\mathrm{d}$ values compared to DESI BAO is compensated by a smaller neutrino mass. Interestingly, the necessary effective mass to obtain this match agrees well with the best fit obtained from CMB data alone in \cref{eq:cmb_negnu_bound}.}
  \label{fig:H0rd_discrepancy}
\end{figure*}

The preference for negative effective masses agrees with the profile likelihood analysis of \cref{sec:profile_likelihood}, but the tension is markedly stronger for the Bayesian analysis presented here. This is not unexpected, as the profile curves are extrapolated from the positive region, which becomes less accurate the further the minimum is in the negative region. Moreover, the profiles are not exact parabolas. A similar effect was seen for Gaussian extrapolations of the Bayesian posterior distribution \cite{Elbers_24,Naredo-Tuero24}.

\cref{fig:negative_mnu_results} also shows the posterior for CMB data alone, which yield
\begin{flalign}
\begin{aligned}
    &\qquad \text{CMB:} \\
    &\qquad \sum m_{\nu,\mathrm{eff}} = -0.11^{+0.12}_{-0.14}\,\si{\eV} \quad (68\%). \label{eq:cmb_negnu_bound}
\end{aligned}
&& 
\end{flalign}

\noindent
This agrees well with \cref{eq:baseline_negnu} and, while showing a preference for negative values, is still compatible with neutrino oscillations to within $2\sigma$. For the first time, we also obtain results for a full-shape power spectrum analysis with effective neutrino masses. Using DESI DR1 (FS+BAO), along with a BBN prior on $\omega_\mathrm{b}$ and CMB priors on the parameters $\theta_*$ and $n_\mathrm{s}$, we find
\begin{flalign}
\begin{aligned}
    &\;\; \text{DESI DR1 (FS+BAO) + BBN + $(\theta_*$, $n_\mathrm{s})_\mathrm{CMB}$:} \\
    &\;\; \sum m_{\nu,\mathrm{eff}} = -0.19^{+0.11}_{-0.16}\,\si{\eV} \quad (68\%). 
\end{aligned}
&& 
\end{flalign}

This particular combination of data and priors exploits the ability of the DESI full-shape analysis to constrain the sum of neutrino masses through the scale-dependent suppression of the power spectrum due to neutrino free streaming, as will be discussed further in \cref{sec:full_shape_sources}. Interestingly, despite relying on different physical signatures and using no CMB information on $\omega_\mathrm{b}$ or $\omega_\mathrm{cdm}$, this constraint also agrees well with \cref{eq:baseline_negnu} and again shows a preference for negative values, while still being compatible with neutrino oscillations to just within $2\sigma$.

\subsubsection{Origin of the anomaly}

We can trace the origin of the preference for negative effective neutrino masses from DESI BAO and CMB to a number of different effects \cite{Elbers_24}. A major contribution comes from the overall amplitude of the BAO distance measurements, which are smaller than what is allowed by CMB data for a $\Lambda$CDM model with positive neutrino masses, as shown in the left panel of \cref{fig:H0rd_discrepancy}. This tension can be expressed as a discrepancy between the values of $H_0r_\mathrm{d}$ obtained from DESI BAO and CMB data \cite{Noriega:2024lzo}. For a fixed $\theta_*$, a small $H_0 r_\mathrm{d}$ from the CMB can be compensated by lowering the total neutrino mass. To illustrate this effect, the right panel of \cref{fig:H0rd_discrepancy} presents the $H_0 r_\mathrm{d}$ posterior distributions for fixed neutrino mass values of $\sum m_{\nu,\mathrm{eff}} = -0.11,\, 0,\, 0.06,$ and $\SI{0.1}{\eV}$ from CMB data alone, showing that the inferred small neutrino mass is associated with the aforementioned discrepancy in $H_0 r_\mathrm{d}$. The shifts in $H_0r_\mathrm{d}$ are almost entirely due to changes in $H_0$ along the geometric degeneracy.

When $\sum m_{\nu,\mathrm{eff}}<0$ is allowed, the effective neutrino mass sum, \cref{eq:cmb_negnu_bound}, preferred by the CMB, independent of any BAO information, yields values of $H_0r_\mathrm{d}$ that agree very well the DESI DR2 BAO measurements. In this case with $\sum m_{\nu,\mathrm{eff}}=\SI{-0.11}{\eV}$ fixed, the CMB constrains $H_0 = 68.96\pm \SI{0.49}{\km\per\s\per\Mpc}$, in line with the distance ladder measurements of \cite{Freedman24}.

\begin{figure*}
    \centering
    \resizebox{\linewidth}{!}{
        \includegraphics[height=0.305\linewidth]{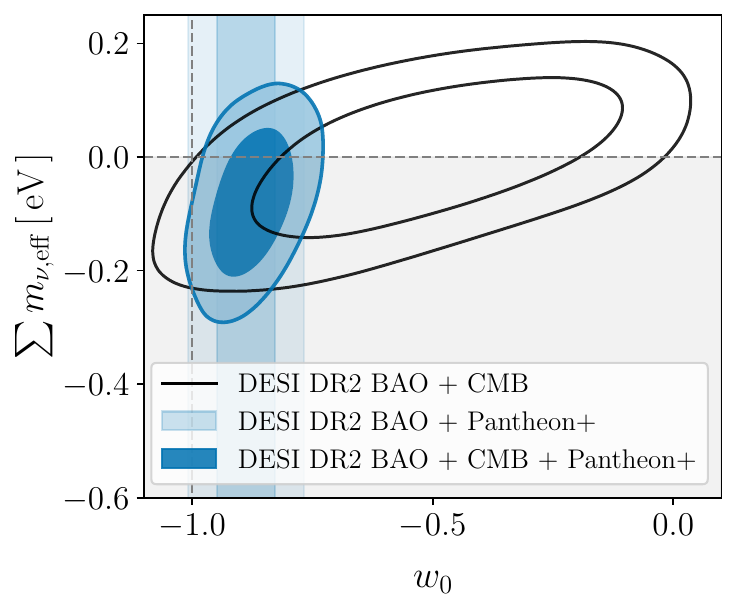}
        \includegraphics[height=0.305\linewidth]{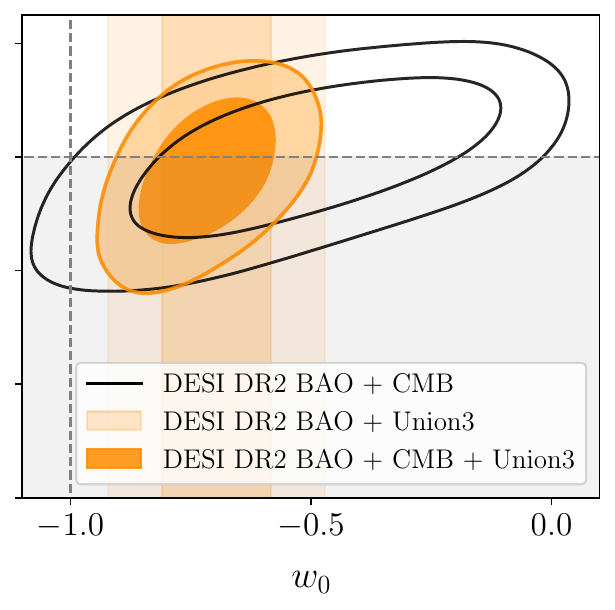}
        \includegraphics[height=0.305\linewidth]{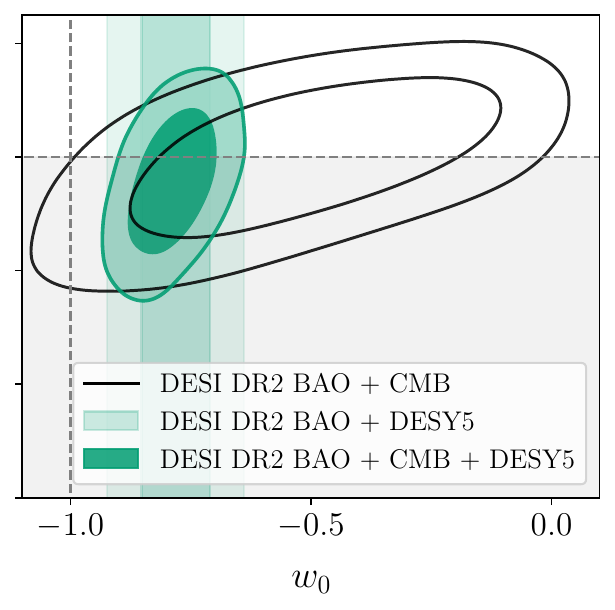}
    }
    \caption{Constraints for $w_0w_a$CDM on the effective neutrino mass parameter, $\sum m_{\nu,\text{eff}}$, and the present-day value of the dark energy equation of state, $w_0$, from DESI DR2 BAO combined with CMB data and three different supernova datasets as indicated. The contours correspond to the 68\% and 95\% posterior regions. The figures demonstrate that one can recover positive neutrino masses in an evolving dark energy model, but that the preference for negative effective neutrino masses depends on how far the DESI + SNe combination pulls away from $\Lambda$CDM (with $w_0=-1$ shown as a vertical dashed line).}
    \label{fig:w0wa_mnu_degeneracy}
\end{figure*}

The neutrino mass tension is also related to the presence of an oscillatory feature in the small-scale CMB temperature power spectrum, which is unaccounted for in $\Lambda$CDM models with positive neutrino masses, and is also degenerate with the $A_\mathrm{lens}$ parameter discussed in \cref{sec:cmb_likelihoods} (see \cite{McCarthy18,Choudhury20,DiValentino20,Allali24,Craig24,Elbers_24,Naredo-Tuero24}). The preference for $A_\mathrm{lens}>1$ is present to different degrees in alternative CMB analyses. We therefore derive constraints on $\sum m_{\nu,\mathrm{eff}}$ for the alternative \texttt{plik} and \texttt{L-H} CMB likelihoods. In both cases, we obtain results that are quite similar to \cref{eq:baseline_negnu}, as shown in \cref{tab:neg_neutrino_constraints}. As expected, the tension is weaker with \texttt{L-H}, which prefers a slightly lower value of $A_\mathrm{lens}$ that is consistent with unity, but the finding of negative effective neutrino masses is clearly robust to the choice of CMB likelihood; however, see also \cite{Allali24,Naredo-Tuero24}. 

A third contribution comes from the large-scale polarization measurements by \emph{Planck}. These data are primarily responsible for constraining the reionization optical depth, $\tau$. For negative effective neutrino masses, CMB data allow smaller values of $\tau$, which improves the fit with \emph{Planck} polarization at large scales. This effect is also related to the $A_\mathrm{lens}$ problem, given that an increase in $\tau$ leads to larger primordial and lensing amplitudes (since $A_\mathrm{s}e^{-2\tau}$ is measured  precisely). Compared to \emph{Planck} \cite{PlanckCosmology2020,Planck20NPIPE}, WMAP found significantly larger values of $\tau$ \cite{Hinshaw13}, which would help to accommodate larger neutrino masses \cite{Craig24,Loverde24}.

We tested explicitly that adopting larger values of $\tau$ shifts the posterior distribution of $\sum m_{\nu,\mathrm{eff}}$ in the positive direction. When fixing the optical depth at $\tau=0.067$ or $\tau=0.074$, corresponding to a $\sim2\sigma$ or $3\sigma$ shift from the baseline value of $\tau=0.054\pm 0.007$, the posteriors move in the positive direction, but the discrepancy with neutrino oscillations remains at $2.3\sigma$ or $1.7\sigma$, respectively. When combined with astrophysical constraints on reionization, such large values of $\tau$ are further disfavored \cite{Paoletti24}. Although recent observations with  JWST might challenge the standard reionization picture \cite{Munoz24}, the \emph{Planck} value of $\tau$ is generally consistent with most astrophysical observations suggesting a later end to reionization \cite{Becker15,Bouwens15,Planck16,Kulkarni19,Choudhury21,Zhu24}. The combination with DESI BAO also limits our ability to explain negative effective neutrino masses in terms of $\tau$ alone, since at fixed neutrino mass, increasing $\tau$ only slightly increases the preferred CMB value of $H_0r_\mathrm{d}$. Nevertheless, if we exclude large-scale CMB polarization data on account of systematic errors, the tension could be explained as a $\sim2\sigma$ statistical fluctuation.

\begin{figure*}
    \centering
    \resizebox{\linewidth}{!}{
        \includegraphics[height=0.365\linewidth]{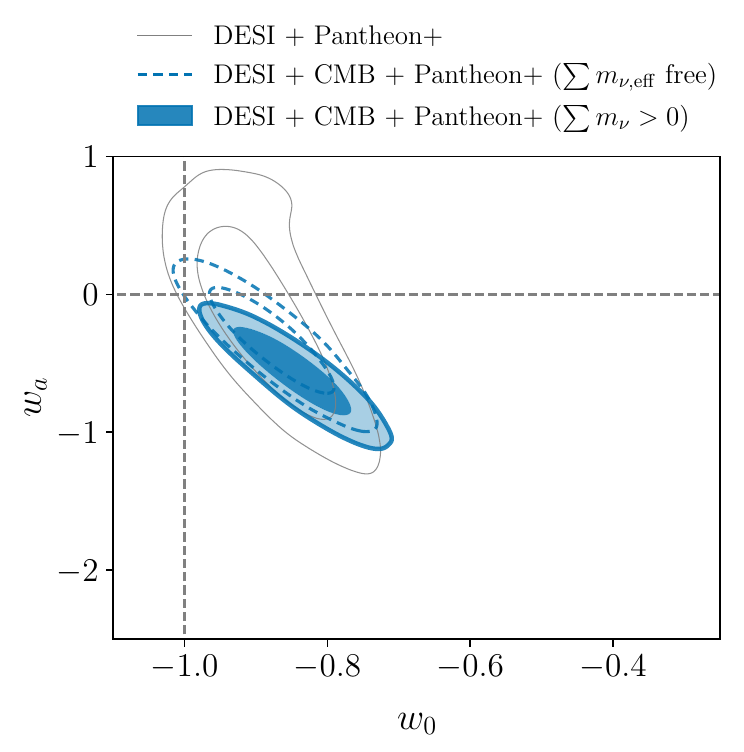}
        \includegraphics[height=0.365\linewidth]{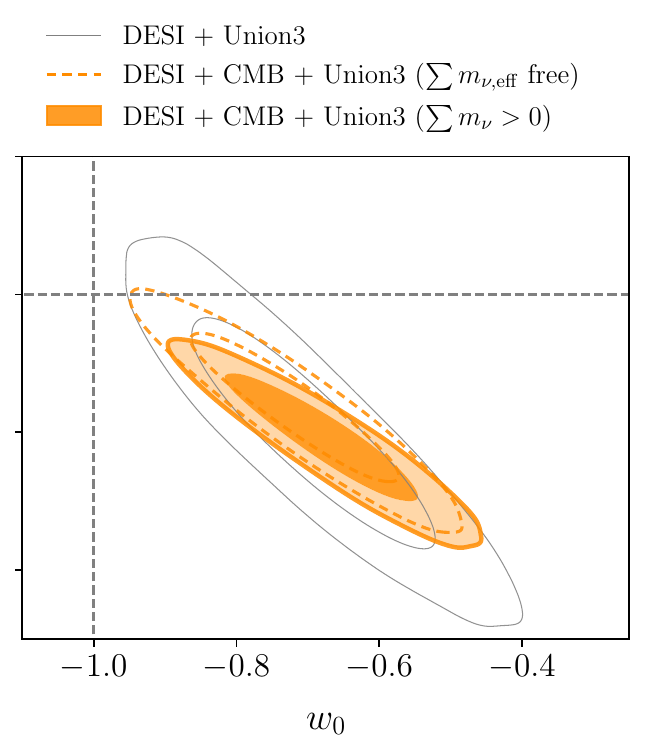}
        \includegraphics[height=0.365\linewidth]{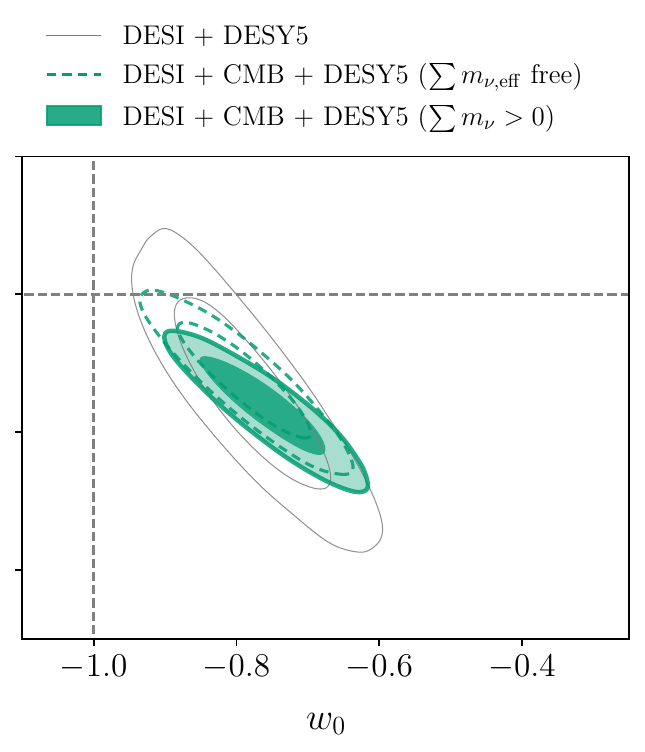}
    }
    \caption{Constraints in the $w_0-w_a$ plane from DESI and SNe, which do not constrain the sum of neutrino masses on their own (hence $\sum m_\nu=\SI{0.06}{\eV}$ is fixed). The plots also show the constraints from DESI, CMB, and SNe Ia data, when $\sum m_{\nu,\mathrm{eff}}$ is free with a broad uniform prior or when $\sum m_\nu$ is free with a $\sum m_\nu>0$ prior. Allowing for negative effective neutrino masses extends the contours closer to $\Lambda$CDM, but does not relax the limits compared to DESI and SNe Ia with neutrino mass fixed.}
    \label{fig:w0wa_sn_results}
\end{figure*}

\subsubsection{Relation to dark energy}

The same effects that are responsible for negative values of $\sum m_{\nu,\mathrm{eff}}$ in $\Lambda$CDM also drive DESI and CMB data away from $\Lambda$CDM toward evolving dark energy models such as $w_0w_a$CDM and mirage dark energy \cite{Elbers_24}. When the background cosmology is generalized to $w_0w_a$CDM, DESI BAO and CMB data can be reconciled with positive neutrino masses. For this combination, we find  
\begin{flalign}
\begin{aligned}
    &\qquad \text{$w_0w_a$CDM: DESI DR2 BAO + CMB:} \\
    &\qquad \sum m_{\nu,\mathrm{eff}} = 0.000^{+0.10}_{-0.081}\,\si{\eV}, \quad (68\%) 
\end{aligned}
&&
\end{flalign}

\noindent
which is fully compatible with the lower limits on the sum of the neutrino masses from neutrino oscillations. Furthermore, the right panel of \cref{fig:negative_mnu_results} shows that the marginalized one-dimensional posterior peaks at a positive value. However, as these data do not strongly constrain the dark energy equation of state parameters, $w_0$ and $w_a$, the results still depend on the priors on $w_0$ and $w_a$. The addition of SNe Ia helps to constrain these parameters. In the case of DESY5, we obtain:
\begin{flalign}
\begin{aligned}
    & \!\text{$w_0w_a$CDM: DESI DR2 BAO + CMB + DESY5:} \\
    & \!\sum m_{\nu,\mathrm{eff}} = -0.044\pm 0.084\,\si{\eV}, \quad (68\%) 
\end{aligned}
&&
\end{flalign}

\noindent
which is also compatible with neutrino oscillations. While the $1\sigma$ error increases by about $\SI{0.03}{\eV}$ compared to the $\Lambda$CDM constraint of \cref{eq:baseline_negnu}, the reduction in tension is mostly due to a shift of $\sim\SI{0.06}{\eV}$ in the central value toward zero.

Compared to the case where $\sum m_\nu=\SI{0.06}{\eV}$ is fixed, allowing $\sum m_\nu>0$ to be free does not significantly impact the constraints on $w_0$ and $w_a$. This is no longer the case with $\sum m_{\nu,\mathrm{eff}}$ free. We find 
\begin{flalign}
\begin{aligned}
    & \!\text{$w_0w_a$CDM: DESI DR2 BAO + CMB + DESY5:} \\
    & \!\left\{
    \begin{aligned}
    \begin{split}
        w_0 &= -0.787\pm 0.061 \\
        w_a &= -0.63^{+0.29}_{-0.26}
    \end{split} \quad (68\%),
    \end{aligned}
    \right.
\end{aligned}
&&
\end{flalign}

\noindent
which corresponds to a $0.6-0.9\sigma$ shift toward $\Lambda$CDM compared to the case with $\sum m_\nu=\SI{0.06}{\eV}$ fixed. 
When we use Pantheon+ or Union3, the conclusions are qualitatively the same, albeit with minor differences in the preferred values, as can be seen from the right panel of \cref{fig:negative_mnu_results} and \cref{tab:neg_neutrino_constraints}. 
In all cases, we have viable solutions with positive neutrino masses when adopting $w_0w_a$CDM. At the same time, the preference for $w_0w_a$CDM is significantly reduced when $\sum m_{\nu,\mathrm{eff}}<0$ is allowed.

The connection with evolving dark energy can be traced to the degeneracy between the dark energy equation of state, $w(z)$, and $\sum m_\nu$ \cite{Hannestad05,Lorenz17,Choudhury18,Vagnozzi18,Upadhye19,Liu20,Choudhury20,Sharma22}, associated primarily with their effects on the expansion history. This degeneracy becomes particularly clear in terms of $\sum m_{\nu,\mathrm{eff}}$, as shown in \cref{fig:w0wa_mnu_degeneracy}. The figures demonstrate that one can obtain positive neutrino masses for all three supernova datasets if one adopts an evolving dark energy model. At the same time, it also shows that imposing $\Lambda$CDM (by fixing $w_0=-1$ and $w_a=0$) forces the effective neutrino mass parameter to be negative.

Although adopting $\sum m_{\nu,\mathrm{eff}}<0$ allows DESI + CMB to be more compatible with $\Lambda$CDM, the preference for $w_0w_a$CDM is not reduced below the level obtained from DESI + SNe with positive neutrino masses, as shown in \cref{fig:w0wa_sn_results}. The reason for this is simply that it is the total matter density fraction, $\Omega_\mathrm{m}=\Omega_\mathrm{cb}+\Omega_\nu$, that determines the late-time expansion history (as probed by BAO, SNe, and the distance to the surface of last scattering), while $\Omega_\mathrm{cb}$ matters for the primary CMB anisotropies. Hence, negative effective neutrino masses may ameliorate the tension between DESI and CMB data, but make no impact on the DESI + SNe constraints.

Interestingly, \cref{fig:w0wa_sn_results} also shows that models with $\sum m_{\nu,\mathrm{eff}}<0$ are compatible with $w_a=0$ and $w_0>-1$, whereas models with $\sum m_\nu>0$ are not. Indeed, when considering the $w$CDM model with a constant equation of state of dark energy, $w$, we find that the data prefer $w>-1$ and $\sum m_{\nu,\mathrm{eff}}<0$, particularly for the combinations with Union3 and DESY5. In this case, the neutrino mass tension can be as high as $3.5\sigma$ for DESI + CMB + DESY5. Refer to \cref{tab:neg_neutrino_constraints} for the parameter constraints for different data combinations.

\begin{figure*}
    \centering
    \resizebox{\linewidth}{!}{
        \includegraphics[width=0.45\linewidth]{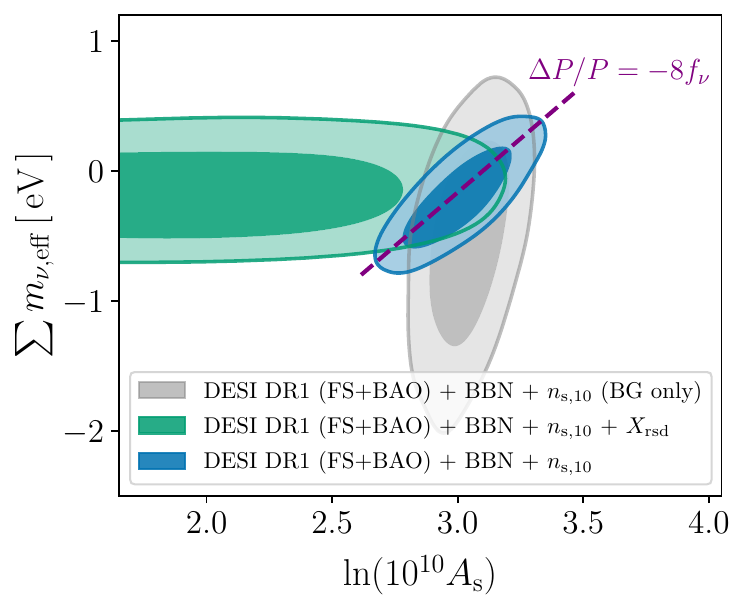}
        \includegraphics[width=0.45\linewidth]{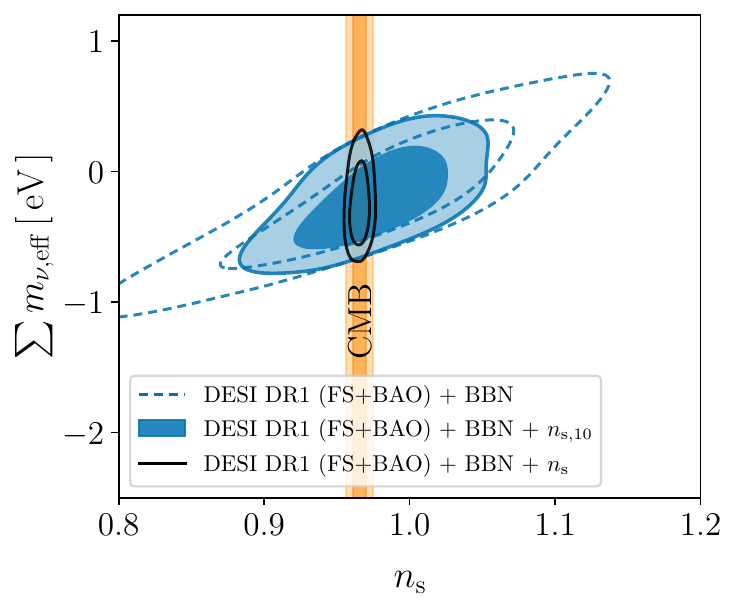}
    }
    \caption{Constraints on the sum of effective neutrino masses, $\sum m_{\nu,\mathrm{eff}}$, and the primordial power spectrum amplitude, $A_\mathrm{s}$, on the left panel and its slope, $n_\mathrm{s}$, on the right panel, assuming the $\Lambda$CDM model. The scaling expected from linear theory, $\Delta P/P=-8f_\nu$, is shown as a purple dashed line. We also show constraints when the theory calculation is modified such that neutrino masses only affect the background expansion (BG only). The figures demonstrate that neutrino mass limits are unaffected when the amplitude becomes unconstrained (by integrating out RSD information through marginalization over the $X_\mathrm{rsd}$ parameter), while the limits become significantly weaker when the slope becomes unconstrained (by removing the weak $n_{\mathrm{s},10}$ prior) or when we only account for background effects. This suggests that, in this mass range, information on $\sum m_\nu$ derives from the free-streaming effect on the shape of the power spectrum rather than from the amplitude or background effects.}
    \label{fig:sources_fullshape2}
\end{figure*}

\subsection{Tension metrics}\label{sec:tension_metrics}

In the previous subsection, we identified a moderate tension between our cosmological constraints on $\sum m_{\nu,\mathrm{eff}}$ and the lower limit from terrestrial oscillation experiments, which applies in the case of three positive neutrino masses under the normal ordering, $\sum m_\nu>\SI{0.059}{\eV}$ \cite{Esteban:2024eli,Capozzi_21,Salas_21}. The tension was quantified by considering the 1D marginalized posterior distribution of $\sum m_{\nu,\mathrm{eff}}$ and computing the probability to exceed (PTE) this lower limit. Here, we briefly consider two alternative tension metrics, following \cite{Gariazzo23,Jiang25}.

The first of these is a goodness-of-fit-loss metric based on the $\Delta\chi^2_\mathrm{MAP}$, evaluated at the maximum a posteriori (MAP) points of our MCMC chains with $\sum m_{\nu,\mathrm{eff}}$ free and with $\sum m_{\nu}=\SI{0.06}{\eV}$ fixed. Here, $\chi^2=-2\ln\mathcal{L}$ is defined in terms of the likelihood, $\mathcal{L}$. We neglect the $\Delta\chi^2_\mathrm{MAP}$ for the oscillation experiments.

The second is a parameter shift metric that quantifies the probability of a non-zero parameter difference between two datasets \cite{Raveri20,Raveri21}. The posterior probability of a given parameter difference, $\Delta x$, between independent datasets $A$ and $B$ is given by
\begin{align}
    P_\Delta(\Delta x) = \int P_A(x)P_B(x-\Delta x)\mathrm{d}x,
\end{align}

\noindent
where $P_A$ and $P_B$ are the posterior probabilities for datasets $A$ and $B$. The probability that there is a non-zero offset is
\begin{align}
    \Delta\equiv \int_{P_\Delta(\Delta x)>P_\Delta(0)} P_\Delta(\Delta x)\mathrm{d}\Delta x.
\end{align}

\begin{table}[b]

\centering
\resizebox{\columnwidth}{!}{
    \begin{tabular}{lcccc}
    \toprule
    Datasets & $\Delta\chi^2_\mathrm{MAP}$ & $\Sigma_\mathrm{MAP}$ & $\Sigma_\Delta$ & $\Sigma_\mathrm{PTE}$ \\
    \midrule
    CMB & $-2.2$ & $1.5\sigma$  & $2.5\sigma$ & $1.6\sigma$ \\
    DESI+CMB (no lensing) & $-6.4$ & $2.5\sigma$ & $3.6\sigma$ & $2.7\sigma$ \\
    DESI+CMB & $-7.2$ & $2.7\sigma$  & $4.0\sigma$ & $3.0\sigma$ \\
    \bottomrule
    \end{tabular}
}
\caption{The tension between cosmological and terrestrial constraints on the sum of neutrino masses, assuming the $\Lambda$CDM model, quantified using three different metrics: a goodness-of-fit-loss metric based on $\Delta\chi^2_\mathrm{MAP}$, a parameter shift metric, $\Delta$, and a direct probability to exceed (PTE) calculated from the 1D marginalized posterior distribution. We convert the test statistics to a Gaussian $N\sigma$-level of confidence denoted as $\Sigma_\mathrm{X}$ for metric $X$. }
\label{tab:tension_metrics}
\end{table}

\noindent
For the terrestrial dataset, we assume a uniform distribution between the lower limit, $\sum m_\nu>\SI{0.059}{\eV}$, and the upper limit from tritium $\beta$-decay, $\sum m_\nu<\SI{1.2}{\eV}$, set by KATRIN \cite{KATRIN24} and converted to a $95\%$ limit as in \cite{Esteban:2024eli}. We experimented with more complicated parametrizations, taking into account the mass splittings, but these gave broadly consistent answers. For all three metrics, we convert the test statistic to a significance level, $N\sigma$, corresponding to an equivalent two-sided Gaussian probability with standard deviation $\sigma$.

For our baseline DESI + CMB combination, we find a tension of $2.7\sigma$ using the goodness-of-fit-loss metric, which agrees well with the $3.0\sigma$ tension obtained from the PTE, in line with the results of \cite{Elbers_24}. In \cite{Gariazzo23,Jiang25}, it was shown that this metric also agrees well with the Bayesian suspiciousness, which we do not consider here. By contrast, these authors showed that the parameter shift metric indicates significantly stronger tensions. We confirm that expectation here, finding a $4.0\sigma$ tension using the parameter shift metric for DESI + CMB. Overall, this validates the $3\sigma$ tension reported above for the baseline DESI + CMB combination. See \cref{tab:tension_metrics} for tensions computed with other data combinations.

\begin{table*}

\centering
\resizebox{\linewidth}{!}{
    \begin{tabular}{lccccc}
    \toprule
    Model/Dataset & $\Omega_\mathrm{m}$ & $H_0$ [km s$^{-1}$ Mpc$^{-1}$] & $\sum m_{\nu,\mathrm{eff}}$ [eV] & $w$ or $w_0$ & $w_a$ \\
    \midrule
    $\mathbf{\Lambda}$\textbf{CDM+}$\mathbf{\sum m_{\nu,\mathrm{eff}}}$ &  &  & &  &  \\
    DESI BAO+CMB (Baseline) & $0.2953\pm 0.0043$ & $68.92\pm 0.38$ & $-0.101^{+0.047}_{-0.056}$ & --- & --- \\
    DESI BAO+CMB (\texttt{plik}) & $0.2948\pm 0.0043$ & $69.06\pm 0.39$ & $-0.099^{+0.050}_{-0.061}$ & --- & --- \\
    DESI BAO+CMB (\texttt{L-H}) & $0.2953\pm 0.0044$ & $68.89\pm 0.39$ & $-0.067^{+0.054}_{-0.064}$ & --- & --- \\
    \hline
    $\mathbf{w}$\textbf{CDM+}$\mathbf{\sum m_{\nu,\mathrm{eff}}}$ &  &  & &  &  \\
    DESI BAO+CMB & $0.2992\pm 0.0075$ & $68.34\pm 0.98$ & $-0.135^{+0.067}_{-0.076}$ & $-0.970\pm 0.047$ & ---\\
    DESI BAO+CMB+Pantheon+ & $0.3033\pm 0.0050$ & $67.76\pm 0.58$ & $-0.161\pm 0.061$ & $-0.942\pm 0.027$ & ---\\
    DESI BAO+CMB+Union3 & $0.3060\pm 0.0057$ & $67.39\pm 0.68$ & $-0.180\pm 0.064$ & $-0.924\pm 0.032$ & ---\\
    DESI BAO+CMB+DESY5 & $0.3074\pm 0.0049$ & $67.19\pm 0.55$ & $-0.191\pm 0.061$ & $-0.914\pm 0.026$ & ---\\
    \hline
    $\mathbf{w_0w_a}$\textbf{CDM+}$\mathbf{\sum m_{\nu,\mathrm{eff}}}$ &  &  & &  &  \\
    DESI BAO+CMB & $0.343\pm 0.025$ & $64.4^{+2.0}_{-2.5}$ & $0.000^{+0.10}_{-0.081}$ & $-0.52\pm 0.25$ & $-1.40\pm 0.76$ \\
    DESI BAO+CMB+Pantheon+ & $0.3074\pm 0.0061$ & $67.62\pm 0.60$ & $-0.078\pm 0.086$ & $-0.876\pm 0.058$ & $-0.34^{+0.27}_{-0.23}$ \\
    DESI BAO+CMB+Union3 & $0.3232\pm 0.0095$ & $66.14\pm 0.87$ & $-0.026\pm 0.084$ & $-0.716\pm 0.096$ & $-0.83\pm 0.36$ \\
    DESI BAO+CMB+DESY5 & $0.3158\pm 0.0062$ & $66.84\pm 0.57$ & $-0.044\pm 0.084$ & $-0.787\pm 0.061$ & $-0.63^{+0.29}_{-0.26}$ \\
    \bottomrule
    \end{tabular}
}
\caption{Constraints on models with the effective neutrino mass parameter, $\sum m_{\nu,\mathrm{eff}}$, for different data combinations.  We report $68\%$ limits for all parameters.}
\label{tab:neg_neutrino_constraints}
\end{table*}

\section{Constraints from free streaming}\label{sec:full_shape_sources}

\noindent
In \cite{DESI2024.VII.KP7B}, DESI presented a constraint from DR1, including both FS and BAO, and using minimal CMB information. Using a BBN prior on $\omega_\mathrm{b}$ and a weak prior on the spectral index, $n_\mathrm{s,10}$, corresponding to ten times the uncertainty of \emph{Planck}, they derived a constraint of $\sum m_\nu<\SI{0.409}{\eV}$ (95\%) in the presence of a cosmological constant.

A question that was left open was the actual source of the information. We will address that here. The left panel of \cref{fig:sources_fullshape2} shows the constraints from this data combination in the plane of $\sum m_\nu$ and the large-scale amplitude of the primordial power spectrum, $\ln(10^{10}A_\mathrm{s})$. The two parameters are degenerate with a slope, $\mathrm{d}\ln(A_\mathrm{s})/\mathrm{d} M_\nu\propto -(\Delta P_\mathrm{m}/P_\mathrm{m})/\Delta f_\nu\approx-8$, which would be naively expected if DESI were constraining neutrino masses through the neutrino free-streaming effect 
on the small-scale amplitude of the power spectrum, but this should be tested.

To gain physical insight into the mechanism behind the constraint, we artificially modify our theory calculations such that neutrino masses only affect the background expansion, but are neglected at the level of cosmological perturbations \cite{Bertolez24}. The remaining effect on the matter power spectrum of raising $\sum m_\nu$, at a fixed cold dark matter density, is a nearly scale-independent suppression combined with a change in the BAO. The constraints obtained with this theoretical modification are significantly weaker, as shown in the left panel of \cref{fig:sources_fullshape2}. This is a clear indication that the FS analysis exploits the scale-dependent effect due to neutrino free streaming.

\begin{figure*}
    \centering
    \resizebox{\linewidth}{!}{
        \includegraphics[width=0.45\linewidth]{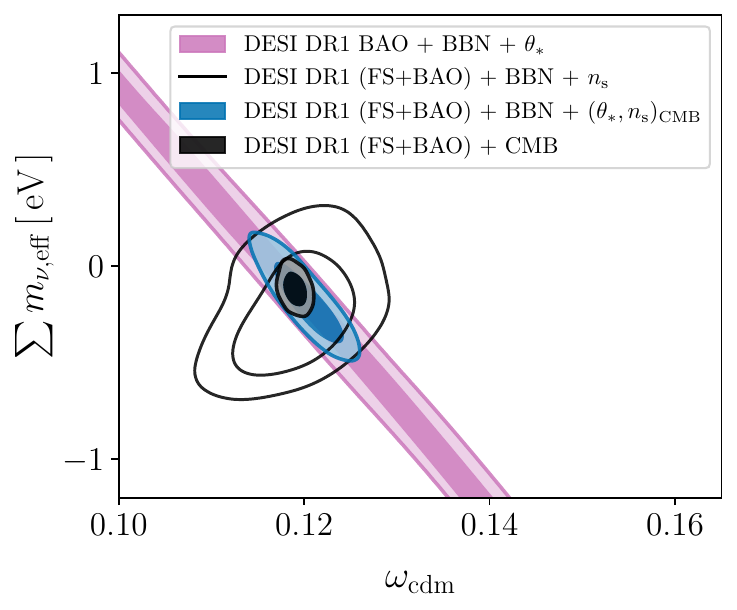}
        \includegraphics[width=0.45\linewidth]{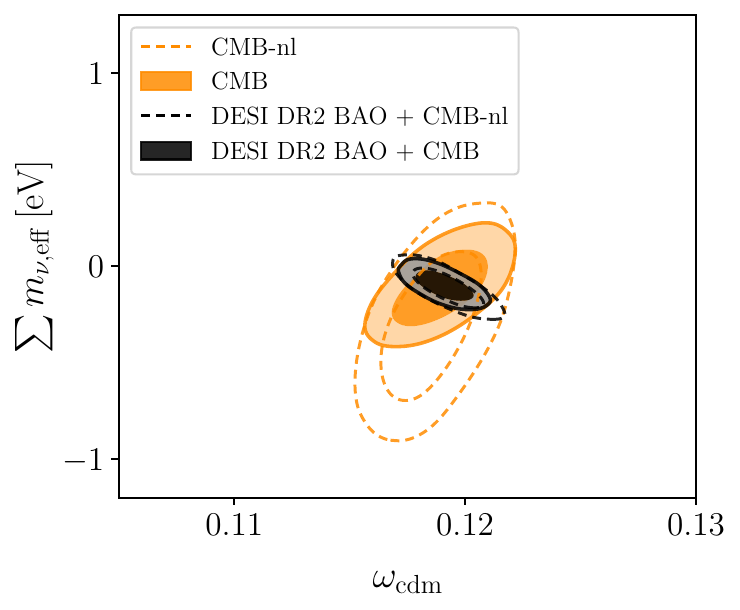}
    }
    \caption{Left: constraints on the sum of effective neutrino masses, $\sum m_{\nu,\mathrm{eff}}$, and $\omega_\mathrm{cdm}$ for data combinations exploiting different signatures of massive neutrinos. Right: constraints on the same parameters from CMB without lensing (CMB-nl) and with lensing (CMB) alone and in combination with DESI DR2 BAO. The lensing information significantly tightens the constraints and helps exclude large negative effective masses.}
    \label{fig:lensing_information}
\end{figure*}

As galaxies are biased tracers of the matter field, the amplitude of the matter power spectrum, $\sigma_8$, is degenerate with the galaxy bias. This degeneracy can be broken by exploiting redshift space distortion (RSD) information on the growth rate. To determine whether the information on neutrino masses is indeed coming from the amplitude, we marginalize over the growth rate, such that the amplitude can no longer be constrained.

To see this, consider that the galaxy power spectrum in redshift space, $P_s(k,\mu)$, can be written in terms of the auto and cross power spectra of the density fluctuation, $\delta$, and the velocity divergence, $\theta$.\footnote{We define $\theta = -\nabla\cdot\mathbf{v} / a H$, with $\mathbf{v}$ the peculiar velocity field. For clarity, we do not include the growth factor in the definition, unlike the common practice.} 
In linear theory, these quantities are related as
\begin{align}
    \theta(\mathbf{k}) = f(k)\delta(\mathbf{k}),
\end{align}

\noindent
where $f(k)$ is the linear growth rate, which is scale dependent in the presence of massive neutrinos. Let us now introduce a new fudge factor, $X_\mathrm{rsd}$, such that the above relationship becomes
\begin{align}
    \theta(\mathbf{k}) = X_\mathrm{rsd}f(k)\delta(\mathbf{k}).
\end{align}

\noindent
By marginalizing over the unknown parameter, $X_\mathrm{rsd}$, we break the theoretical link between $\theta$ and $\delta$ that allows $\sigma_8$ to be constrained. The left panel of \cref{fig:sources_fullshape2} shows that, although the amplitude indeed becomes unconstrained in this case, the limits on $\sum m_\nu$ are not significantly affected. This suggests that the constraint on $\sum m_\nu$ is unrelated to the amplitude. 
In principle, information could still be coming from the slight differences in the suppression of the amplitude at different redshifts. However, this effect is negligible.

Alternatively, the free-streaming effect could be detected from the shape of the power spectrum. \cref{fig:sources_fullshape1} shows that, on the scales used in the FS analysis, the neutrino effect manifests as a change in the slope compared to the power spectrum on large scales. The analysis must then be anchored to an external prior on the large-scale slope, $n_\mathrm{s}$, from the CMB. The right panel of \cref{fig:sources_fullshape2} shows what happens when the weak prior on $n_\mathrm{s}$ is replaced with a less informative uniform prior on $n_\mathrm{s}\in[0.8,1.2]$: the neutrino mass limits degrade significantly. Conversely, imposing a stronger prior on $n_\mathrm{s}$, corresponding to the $1\sigma$ uncertainty from \emph{Planck}, yields \cite{DESI2024.VII.KP7B}
\begin{flalign}
\begin{aligned}
    &\qquad \text{DESI DR1 (FS+BAO) + BBN + $n_{\mathrm{s}}$:} \\
    &\qquad \, \sum m_\nu < 0.300\,\si{\eV} \quad  (95\%), 
\end{aligned}
&& \label{eq:fs_ns_limit}
\end{flalign}

\noindent
assuming $\Lambda$CDM. These tests suggest that DESI alone, with minimal external information, is sensitive to the free-streaming effect through the shape of the power spectrum rather than its amplitude or through pure background effects, in line with the findings of \cite{Brieden22}. 

Although \cref{eq:fs_ns_limit} is not competitive with the limits obtained from the geometrical effect, this nevertheless represents an interesting limit that does not rely on any CMB determination of the physical density, $\omega_\mathrm{cb}$. Continuing along these lines, it is worth obtaining a constraint from DESI that exploits both geometrical and shape information, without using any CMB prior on $\omega_\mathrm{cdm}$ or $\omega_\mathrm{b}$. For the combination DESI DR1 (FS+BAO) + BBN with CMB priors on $\theta_*$ and $n_\mathrm{s}$, we find:
\begin{flalign}
\begin{aligned}
    &\;\; \text{DESI DR1 (FS+BAO) + BBN + ($\theta_*$, $n_{\mathrm{s}})_\mathrm{CMB}$:} \\
    &\;\; \, \sum m_\nu < 0.193\,\si{\eV} \quad  (95\%). 
\end{aligned}
&& \label{eq:fs_thetas_ns_limit}
\end{flalign}

\noindent
We already considered this combination in \cref{sec:effective_neutrinos}, where we showed that it preferred negative effective neutrino masses, but was still compatible with the lower limit under the normal mass ordering to just within $2\sigma$. Interestingly, the geometric constraint from DESI BAO + $\theta_*$ and the constraint based on shape information from DESI (FS+BAO) + $n_\mathrm{s}$ are highly complementary, as can be seen from the left panel of \cref{fig:lensing_information}. Physically, this can be traced to the fact that CDM and neutrinos are essentially indistinguishable as far as the late-time expansion history is concerned, but affect differently the formation of structure. We can use the observed information, $I(\theta)=-\mathrm{d}^2\log p(\theta\rvert y)/\mathrm{d}\theta^2$, or the second derivative of the posterior probability with respect to the parameter $\theta$, as a criterion to quantify the information content on $\theta=\sum m_\nu$ of different data combinations. In terms of this quantity, we find that geometry and shape each contribute about half of the observed information to the limit in \cref{eq:fs_thetas_ns_limit}.

However, it has been shown that projection effects significantly distort the geometrical information extracted from the full-shape analysis alone, even with a CMB prior on $\theta_*$ \cite{Noriega:2024lzo}. Furthermore, when additional information from the CMB is incorporated, particularly on $\omega_\mathrm{b}$ and $\omega_\mathrm{cdm}$, the constraints on the neutrino mass become small to the point that the suppression of the power spectrum is unobservable with current sensitivity.

As another dataset that probes large-scale density perturbations, CMB lensing is also sensitive to neutrino free streaming. In the right panel of \cref{fig:lensing_information}, we present constraints from the CMB with and without CMB lensing. We find that lensing information particularly helps to exclude the tail of the distribution with large negative values of $\sum m_{\nu,\mathrm{eff}}$, as well as slightly improving the precision overall. When combined with DESI DR2 BAO, lensing information particularly helps to constrain $\omega_\mathrm{cdm}$ without shifting the posterior in terms of $\sum m_{\nu,\mathrm{eff}}$.

\section{Discussion and conclusions}\label{sec:conclusions}

In recent years, cosmological upper limits on the sum of neutrino masses have steadily crept down and approached the lower limits from neutrino oscillations (e.g. \cite{Vagnozzi17,Loureiro19,PlanckCosmology2020,Choudhury20,Palanque20,DiValentino21,Brieden22}), leaving little room for a measurement that satisfies both laboratory and astrophysical constraints, when assuming the standard $\Lambda$CDM model of cosmology.

This trend continues with the latest results from the DESI collaboration, presented in this paper. Using BAO measurements of more than 14 million galaxies and quasars from the second data release (DR2), combined with external CMB data from \emph{Planck} (PR4) and ACT, we have placed our tightest constraints yet on the sum of neutrino masses. We summarize the key results obtained in this paper in \cref{tab:summary_table}. In our baseline setup, assuming the $\Lambda$CDM model and three degenerate neutrino species with the minimal physical prior that $\sum m_\nu>0$, we derive an upper limit of $\sum m_\nu<\SI{0.0642}{\eV}$ (95\%) from the combination of DESI DR2 BAO and CMB with the PR4 \texttt{CamSpec} likelihood \cite{Efstathiou2021}, including CMB lensing from ACT \cite{ACTDR62024}. We also place constraints on the effective number of neutrino species, finding $N_\mathrm{eff}=3.23^{+0.35}_{-0.34}$ (95\%) from the same data combination. This is consistent with the Standard Model prediction that $N_\mathrm{eff}=3.044$ \cite{Froustey20,Bennett21,Drewes24}.

The upper limit on $\sum m_\nu$ in our baseline setup is close to the lower limit from neutrino oscillations, which applies in the case of three neutrino species with positive masses under the normal mass ordering, $\sum m_\nu>\SI{0.059}{\eV}$ \cite{Esteban:2024eli,Capozzi_21,Salas_21}. Moreover, this result already challenges the alternative inverted mass ordering. In an analysis where we account for the fact that a large fraction of the posterior volume violates the constraints for both mass orderings, we nevertheless find a Bayes factor of $K=10$ in favor of the normal mass ordering, when adopting a uniform prior on the lightest neutrino mass, $m_l$, and Gaussian priors on the squared mass splittings based on oscillation constraints \cite{Esteban:2024eli}. In this setup, we constrain the lightest neutrino mass to be $m_l<\SI{0.023}{\eV}$ (95\%).

Alternative choices for the CMB likelihood or the inclusion of SNe data lead to small differences in the derived upper limits. The greatest limit is found in the case where the baseline \texttt{CamSpec} likelihood \cite{Efstathiou2021} is replaced with \texttt{L-H} \cite{Tristram2024}, $\sum m_\nu<\SI{0.0774}{\eV}$ (95\%). The inclusion of SNe data has a comparatively smaller impact, with the greatest limit resulting from the inclusion of DESY5 SNe \cite{DESY5SN2024}, $\sum m_\nu<\SI{0.0744}{\eV}$ (95\%). However, in all cases we find that the marginalized posterior distribution peaks at $\sum m_\nu=\SI{0}{\eV}$, suggesting that there is no evidence from our data for non-zero neutrino masses. Instead, our baseline constraints are affected by prior weight effects, such that the $\sum m_\nu>0$ prior pulls the posterior distribution away from the maximum likelihood value.

Using the profile likelihood method, a frequentist approach that does not rely on explicit priors, we establish that the minimum of the parabolic likelihood curve is in the unphysical negative mass range when extrapolated. In the case of $\Lambda$CDM, we find a 95\% Feldman-Cousins upper limit of $\sum m_\nu<\SI{0.053}{\eV}$, which corrects for the physical lower limit of zero neutrino mass.

We continued our investigations with a more general Bayesian analysis of a model with the effective cosmological neutrino mass parameter, $\sum m_{\nu,\mathrm{eff}}$, of \cite{Elbers_24}. This parameter coincides exactly with $\sum m_\nu$ for positive values, but allows for negative energy densities, which can be seen as a mathematical continuation of $\sum m_\nu$ to negative values. Assuming $\Lambda$CDM, we find $\sum m_{\nu,\mathrm{eff}} = -0.101^{+0.047}_{-0.056}\,\si{\eV}$ (68\%; DESI DR2 + CMB) and a $3.0\sigma$ tension with the lower limit for the normal mass ordering. From CMB data alone, we obtain $\sum m_{\nu,\mathrm{eff}} = -0.11_{-0.14}^{+0.12}\,\si{\eV}$ (68\%). By keeping this value fixed, and fitting the remaining cosmological parameters within the $\Lambda$CDM framework, using again CMB data without other datasets, we find that $H_0 r_\mathrm{d}$ remarkably matches the value derived solely from DESI DR2 BAO, as illustrated in \cref{fig:H0rd_discrepancy}. This again shows consistency of CMB and DESI data within $\Lambda$CDM when we allow for the effective neutrino mass parameter.

In the context of the main DESI results presented in the key paper \cite{DESI.DR2.BAO.cosmo}, this neutrino tension could be seen as a further hint that the $\Lambda$CDM model is being challenged by DESI BAO and CMB data. While the level of statistical significance is by no means conclusive, it is interesting to consider explanations that could reconcile cosmological neutrino mass limits with the laboratory constraints.

\begin{table}[t]
    \resizebox{\columnwidth}{!}{
    \begin{ruledtabular}
    \centering
    \begin{tabular}{p{6.6cm} r}
        \textbf{Result} & \textbf{Section} \\
        \hline
        Strongest limits from DESI BAO and CMB\par $\sum m_\nu<\SI{0.0642}{\eV}$ $(95\%)$  in $\Lambda$CDM\par $\sum m_\nu<\SI{0.163}{\eV}$ $(95\%)$ in $w_0w_a$CDM & \cref{sec:baseline_results} \\
        Limit on the number of relativistic species\par $N_\mathrm{eff}=3.23^{+0.35}_{-0.34}$ $(95\%)$ & \cref{sec:Neff} \\
        Preference for the normal mass ordering and\par limit on the lightest neutrino mass\par $m_l<\SI{0.023}{\eV}$ $(95\%)$ & \cref{sec:neutrino_mass_ordering} \\
        Frequentist Feldman-Cousins limits\par $\sum m_\nu<\SI{0.053}{\eV}$ $(95\%)$ in $\Lambda$CDM\par $\sum m_\nu<\SI{0.177}{\eV}$ $(95\%)$ in $w_0w_a$CDM & \cref{sec:profile_likelihood} \\
        Effective neutrino mass parameter constraints\par $\sum m_{\nu,\mathrm{eff}} = -0.101^{+0.047}_{-0.056}\,\si{\eV}$ $(68\%)$ in $\Lambda$CDM\par $\sum m_{\nu,\mathrm{eff}} = 0.000^{+0.10}_{-0.082}\,\si{\eV}$ $(68\%)$ in $w_0w_a$CDM & \cref{sec:effective_neutrinos}\\
        Tension with the oscillations lower limit\par between $2.7\sigma$ and $4.0\sigma$, assuming $\Lambda$CDM & \cref{sec:tension_metrics}\\
        Strongest constraint from free streaming\par $\sum m_\nu<\SI{0.193}{\eV}$ $(95\%)$ & \cref{sec:full_shape_sources}\\
    \end{tabular}
    \end{ruledtabular}
    }
    \caption{Summary of the key results obtained in this paper and the sections where those results are discussed.}
    \label{tab:summary_table}
\end{table}

First, despite extensive testing \cite{DESI2024.III.KP4,Y3.clust-s1.Andrade.2025,KP4s9-Perez-Fernandez,KP4s10-Mena-Fernandez, KP4s11-Garcia-Quintero}, the possibility of systematic errors cannot be dismissed. Considering the mismatch in $H_0r_\mathrm{d}$ between DESI BAO and CMB, the tension could be resolved by scaling the isotropic distance measurements of all galaxy tracers, but this would require unknown systematic errors in excess of $1\%$, well above the size of all known systematics \cite{Y3.clust-s1.Andrade.2025}. The consistency of DESI DR2 BAO \cite{DESI.DR2.BAO.cosmo} with prior measurements from DESI DR1 \cite{DESI2024.III.KP4}, SDSS \cite{2021MNRAS.500..736B,2020MNRAS.498.2492G,Alam21,DESI.DR2.BAO.cosmo} and supernova data \cite{PantheonPlus2022,Brout:2022,Union32023,DESY5SN2024} makes this less likely. On the CMB side, effects associated with the lensing amplitude, $A_\mathrm{lens}$, and the optical depth, $\tau$, could also play a role, but the consistency between \texttt{L-H} and \texttt{CamSpec} shows that other factors might be needed as well.

Models of evolving dark energy, further elaborated upon in the accompanying paper on dark energy \cite{Y3.cpe-s1.Lodha.2025}, also provide a possible solution. In our baseline setup with DESI BAO and CMB data, we find $\sum m_\nu<\SI{0.163}{\eV}$ (95\%) when assuming a dynamical dark energy with an evolving equation of state, i.e. the $w_0w_a$CDM model. Our frequentist analysis yields an upper limit of $\sum m_\nu<\SI{0.177}{\eV}$ (95\%), which is only slightly higher than the Bayesian limit. Even when allowing for negative effective neutrino masses, we obtain viable solutions with positive neutrino masses when adopting the $w_0w_a$CDM parametrization. This is made possible by the degenerate effects of $\sum m_\nu$ and the dark energy equation of state parameters, $w_0$ and $w_a$, on the cosmic expansion history. The neutrino fluid violates the null-energy condition when $\sum m_{\nu,\mathrm{eff}}<0$, which it has in common with a single-field dark energy component in the phantom regime $(w<-1)$ \cite{Y3.cpe-s1.Lodha.2025}. This may point to a common origin of the observed trends.

Alternatively, the tension with neutrino oscillations could be relaxed by introducing new physics in the neutrino sector, such as neutrino decay \cite{Serpico07,Escudero19,Escudero20,Chacko21,Abellan22}. However, negative effective masses require more exotic explanations, such as long-range interactions \cite{Esteban21,Archidiacono22,Esteban22,Bogorad24,Bottaro24,Craig24,Kaplan25}. New physics at the time of recombination could also provide an explanation \cite{Baryakhtar24,Lynch24,Mirpoorian24} by changing the sound horizon at the baryon drag epoch, $r_\mathrm{d}$, from the $\Lambda$CDM expectation. This may help to reconcile the CMB and BAO determinations of $H_0r_\mathrm{d}$ without invoking changes to the late-time expansion history. To differentiate between these possibilities, it is interesting to obtain constraints on $\sum m_\nu$ that do not rely on calibration of the BAO standard ruler, $r_\mathrm{d}$, but instead exploit the effect of neutrino free streaming on the galaxy power spectrum.

Using a BBN prior on $\omega_\mathrm{b}$ \cite{Schoeneberg:2024} and a weak prior on the spectral index, $n_\mathrm{s}$, corresponding to ten times the uncertainty of \emph{Planck} \cite{PlanckCosmology2020}, DESI already derived such a constraint, $\sum m_\nu<\SI{0.409}{\eV}$ (95\%), from the full-shape clustering of DR1 galaxies and quasars \cite{DESI2024.VII.KP7B}. In this paper, we strengthen the constraint by adding stronger CMB priors on $\theta_*$ and $n_\mathrm{s}$, finding $\sum m_\nu<\SI{0.193}{\eV}$ (95\%). Despite relying on different physical mechanisms, this combination yields $\sum m_{\nu,\mathrm{eff}} = -0.19^{+0.11}_{-0.16}\,\si{\eV}$ (68\%) in terms of the effective neutrino mass parameter, which agrees well with the aforementioned CMB-only and CMB~+~DESI constraints. While not currently competitive with constraints from the expansion history, further improvements can be expected from the full-shape clustering analysis of DR2 tracers.

Our results indicate that there is a neutrino mass tension between the latest cosmological limits on $\sum m_\nu$ from DESI BAO and CMB data, derived within the $\Lambda$CDM framework, and the values inferred from neutrino oscillations. The tension stands at $3\sigma$ using our baseline choice of datasets, which is a slight increase compared to DESI DR1 \cite{DESI2024.VI.KP7A,DESI2024.VII.KP7B,Elbers_24}. Our findings are supported by complementary statistical analyses using frequentist and Bayesian methods, and could point to unidentified systematic errors or an inconsistency with the $\Lambda$CDM model. Upcoming analyses of DESI data, including the full-shape clustering of DR2 galaxies and quasars, will shed further light on this issue. With a projected sensitivity of $\SI{0.02}{\eV}$ \cite{DESI2016a.Science}, the final DESI survey could make a significant detection of neutrino mass or further challenge the $\Lambda$CDM model if the tension persists.

\section{Data Availability}
The data used in this analysis will be made public along the Data Release 2 (details in \url{https://data.desi.lbl.gov/doc/releases/}). BAO likelihoods for the DESI DR2 likelihoods will be provided integrated in the \texttt{cobaya} code at \url{https://github.com/CobayaSampler/bao_data}.

\acknowledgments

This work used the DiRAC@Durham facility managed by the Institute for Computational Cosmology on behalf of the STFC DiRAC HPC Facility (www.dirac.ac.uk). The equipment was funded by BEIS capital funding via STFC capital grants ST/K00042X/1, ST/P002293/1, ST/R002371/1 and ST/S002502/1, Durham University and STFC operations grant ST/R000832/1. DiRAC is part of the National e-Infrastructure. WE, CSF, and AM acknowledge STFC Consolidated Grant ST/X001075/1 and support from the European Research Council (ERC) Advanced Investigator grant DMIDAS (GA 786910).

AA, HEN, DG and GN acknowledge support by CONAHCyT CBF2023-2024-162. AA, DG and HEN acknowledge support by PAPIIT IA101825. AA acknowledges support by PAPIIT IG102123. GN and DG also acknowledge support by DAIP-UG and DCI-UG DataLab.
CGQ acknowledges support provided by NASA through the NASA Hubble Fellowship grant HST-HF2-51554.001-A awarded by the Space Telescope Science Institute, which is operated by the Association of Universities for Research in Astronomy, Inc., for NASA, under contract NAS5-26555.
KN acknowledges support from the Royal Society grant number URF\textbackslash R\textbackslash 231006. MI acknowledges that this material is based upon work supported in part by the Department of Energy, Office of Science, under Award Number DE-SC0022184 and also in part by the U.S. National Science Foundation under grant AST2327245.

This material is based upon work supported by the U.S. Department of Energy (DOE), Office of Science, Office of High-Energy Physics, under Contract No. DE–AC02–05CH11231, and by the National Energy Research Scientific Computing Center, a DOE Office of Science User Facility under the same contract. Additional support for DESI was provided by the U.S. National Science Foundation (NSF), Division of Astronomical Sciences under Contract No. AST-0950945 to the NSF’s National Optical-Infrared Astronomy Research Laboratory; the Science and Technology Facilities Council of the United Kingdom; the Gordon and Betty Moore Foundation; the Heising-Simons Foundation; the French Alternative Energies and Atomic Energy Commission (CEA); the National Council of Humanities, Science and Technology of Mexico (CONAHCYT); the Ministry of Science, Innovation and Universities of Spain (MICIU/AEI/10.13039/501100011033), and by the DESI Member Institutions: \url{https://www.desi.lbl.gov/collaborating-institutions}. Any opinions, findings, and conclusions or recommendations expressed in this material are those of the author(s) and do not necessarily reflect the views of the U. S. National Science Foundation, the U. S. Department of Energy, or any of the listed funding agencies.

The authors are honored to be permitted to conduct scientific research on I'oligam Du'ag (Kitt Peak), a mountain with particular significance to the Tohono O’odham Nation.

\bibliographystyle{mod-apsrev4-2}
\bibliography{main,DESI_supporting_papers}

\appendix

\section{Details of the effective neutrino mass model}\label{sec:effective_mass_details}

The energy density of massive neutrinos is given by
\begin{align}
    \rho_\nu(z) = \sum_{i=1}^{N_\nu}\frac{g_i (1+z)^4}{2\pi^2} \int_0^\infty\frac{p^2 \epsilon(p,m_i)}{1+e^{p/T_{\nu,0}}}\mathrm{d}p, \label{eq:neutrino_energy_density}
\end{align}

\noindent
where $g_i$ is the degeneracy factor and $m_i$ the mass of neutrino species $i$. The neutrino energy is $\epsilon(p,m)=\sqrt{p^2+a^2m^2}$ in terms of the scale factor, $a$, and the present-day neutrino temperature is $T_{\nu,0}$. To describe perturbations in the cosmological neutrino fluid, it is standard to write the neutrino phase-space distribution as $f(\mathbf{x},\mathbf{p},\tau)=\bar{f}(p)\left[1+\Psi(\mathbf{x},\mathbf{p},\tau)\right]$, where $\bar{f}$ is the background Fermi-Dirac distribution and $\Psi$ is the perturbation, which is commonly expressed as a Legendre series in Fourier space \cite{Ma95}
\begin{align}
    \Psi(\mathbf{k},\mathbf{p},\tau) = \sum_{\ell=0}^\infty (-1)^\ell (2\ell+1)\Psi_\ell(\mathbf{k},p,\tau)P_\ell(\hat{k}\cdot\hat{n}),
\end{align}

\noindent
where $\mathbf{p}=p\hat{n}$ is the comoving momentum. In terms of $\Psi$, the Boltzmann equation becomes an infinite tower of equations. In conformal Newtonian gauge, these are given by \cite{Ma95}
\begin{align}
    \dot{\Psi}_0 &= -\frac{pk}{\epsilon}\Psi_1 - \dot{\phi}\frac{\mathrm{d}\ln \bar{f}}{\mathrm{d}\ln p},\\
    \dot{\Psi}_1 &= \frac{pk}{3\epsilon}(\Psi_0 - 2\Psi_2) - \frac{\epsilon k}{3p}\psi\frac{\mathrm{d}\ln \bar{f}}{\mathrm{d}\ln p}, \\
    \dot{\Psi}_\ell &= \frac{pk}{(2\ell+1)\epsilon}\left[\ell\Psi_{\ell-1} - (\ell+1)\Psi_{\ell+1}\right], \;\;\; (\ell\geq2),
\end{align}

\noindent
where dots denote conformal time derivatives, and $\phi$ and $\psi$ are the metric perturbations.

In the effective neutrino mass model of \cite{Elbers_24}, the neutrino energy in the above equations (including in similar expressions as \cref{eq:neutrino_energy_density} for the pressure, shear, etc.) is replaced by an effective neutrino energy, defined as 
\begin{align}
    \epsilon_\mathrm{eff}=\mathrm{sgn}(m_{\nu,\mathrm{eff}})\sqrt{p^2+a^2m_{\nu,\mathrm{eff}}^2},
\end{align}

\noindent
such that the sign is reversed for $m_{\nu,\mathrm{eff}}<0$. This is easily implemented in cosmological Boltzmann solvers and allows the model to be extended to negative values. It was shown that negative values effectively reverse the cosmological effects of massive neutrinos. For instance, the suppression of the matter power spectrum becomes an enhancement, and similarly the effect on cosmic distances is reversed.

We maintain the same expansion history in the early Universe with $N_\mathrm{eff}=3.044$ by adjusting the number of massless neutrinos, $N_\mathrm{ur}$. We choose the standard \texttt{CLASS} value of $N_\mathrm{ur}=0.00441$ for $\sum m_{\nu,\mathrm{eff}}>0$, but set $N_\mathrm{ur}=6.08627$ for $\sum m_{\nu,\mathrm{eff}}<0$.

\begin{figure}[b!]
  \centering
  \resizebox{\linewidth}{!}{
    \includegraphics[width=0.95\linewidth]{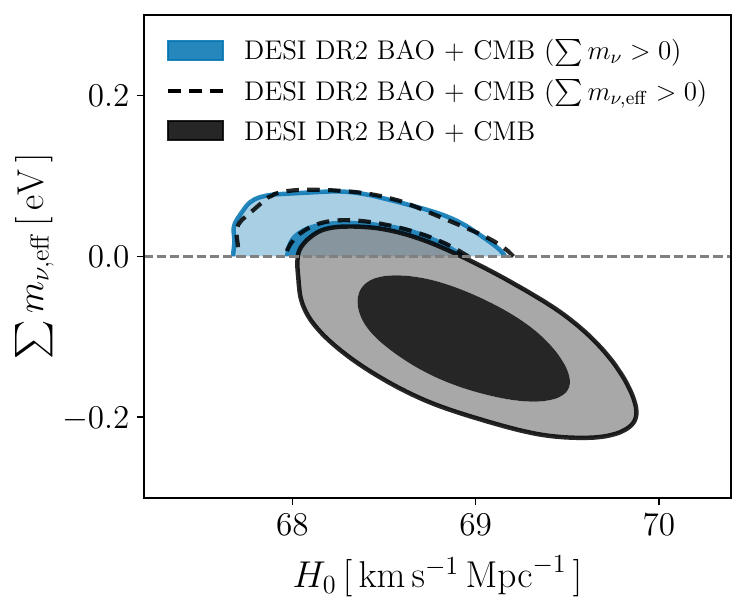}
  }
  \caption{Constraints on $\sum m_{\nu,\mathrm{eff}}$ and $H_0$ from the combination of DESI DR2 BAO with CMB. When $\sum m_{\nu,\mathrm{eff}}$ is restricted to positive values, we recover the results from the standard analysis in terms of $\sum m_\nu$. The figure also shows the projection effect of the $\sum m_\nu>0$ prior on $H_0$. That the 95\% contours of the general case appear to line up with the 68\% contours of the standard case is entirely coincidental.}
  \label{fig:neg_mnu_validation}
\end{figure}

By definition, the model coincides exactly with the standard analysis in terms of $\sum m_\nu$ when the effective neutrino mass parameter is restricted to positive values, as shown in \cref{fig:neg_mnu_validation}.

\vfill

\end{document}